%% file: pim1_beamline.tex
\documentclass[aps,prc,twocolumn,superscriptaddress,nofootinbib]{revtex4-1}

\usepackage{amssymb}
\usepackage{csquotes}
\usepackage[font=singlespacing]{subcaption}
\usepackage{xcolor}
\usepackage{graphicx}
\usepackage{textgreek}
\usepackage{amsmath}
\usepackage[colorlinks=true, linkcolor=black, urlcolor=blue, citecolor=gray]{hyperref}
\usepackage{cleveref}

\usepackage{setspace}

 \usepackage{lineno}

\begin{document}



\title{Characterization of Muon and Electron Beams in the Paul Scherrer Institute PiM1 Channel for the MUSE Experiment}

\author{E.~Cline}
\email{Dr. Ethan Cline (ethan.cline@stonybrook.edu)}
\affiliation{Department of Physics and Astronomy, Stony Brook University, Stony Brook, NY, 11794, USA}
\affiliation{Department of Physics and Astronomy, Rutgers, The State University of New Jersey, Piscataway, NJ, 08855, USA}
\author{W.~Lin}
\affiliation{Department of Physics and Astronomy, Rutgers, The State University of New Jersey, Piscataway, NJ, 08855, USA}
\author{P.~Roy}
\affiliation{{Randall Laboratory of Physics, University of Michigan, Ann Arbor, MI 48109, USA}}
\author{P.~E.~Reimer}
\affiliation{Physics Division, Argonne National Laboratory, Lemont, IL, 60439, USA}
\author{K.~E.~Mesick}
\affiliation{Department of Physics and Astronomy, Rutgers, The State University of New Jersey, Piscataway, NJ, 08855, USA}
\author{A.~Akmal}
\affiliation{Department of Engineering, Physical, and Computer Sciences, Montgomery College, Rockville, MA, 20850, USA}
\author{A.~Alie}
\affiliation{Department of Physics and Astronomy, Rutgers, The State University of New Jersey, Piscataway, NJ, 08855, USA}
\author{H.~Atac}
\affiliation{Department of Physics, Temple University, Philadelphia, PA, 19122, USA}
\author{A.~Atencio}
\affiliation{Department of Physics, Temple University, Philadelphia, PA, 19122, USA}
\author{C.~Ayerbe Gayoso}
\altaffiliation{Current address: Department of Physics and Astronomy, Mississippi State University, Starkville, MS, 39762, USA}
\affiliation{Department of Physics, The College of William \& Mary, Williamsburg, VA, 23187, USA}
\author{N.~Benmouna}
\affiliation{Department of Engineering, Physical, and Computer Sciences, Montgomery College, Rockville, MA, 20850, USA}
\author{F.~Benmokhtar}
\affiliation{Department of Physics, Duquesne University, Pittsburgh, PA, 15282, USA}
\author{J.~C.~Bernauer}
\affiliation{Department of Physics and Astronomy, Stony Brook University, Stony Brook, NY, 11794, USA}
\affiliation{RIKEN BNL Research Center, Brookhaven National Laboratory, Upton, NY, 11973, USA}
\author{W.~J.~Briscoe}
\affiliation{Department of Physics, The George Washington University, Washington, D.C. 20052, USA}
\author{J.~Campbell}
\affiliation{Department of Physics and Atmospheric Science, Dalhousie University, Halifax, NS B3H 4R2, Canada}
\affiliation{Department of Physics and Environmental Science, Saint Mary's University, Halifax, NS B3H 3C3, Canada}
\author{D.~Cohen}
\affiliation{Racah Institute of Physics, The Hebrew University of Jerusalem, Jerusalem, Israel}
\author{E.~O.~Cohen}
\affiliation{School of Physics and Astronomy, Tel Aviv University, Tel Aviv, 69978, Israel}
\author{C.~Collicott}
\affiliation{Department of Physics, The George Washington University, Washington, D.C. 20052, USA}
\author{K.~Deiters}
\affiliation{Paul Scherrer Institute, Villigen, CH-5232, Switzerland}
\author{S.~Dogra}
\affiliation{Department of Physics and Astronomy, Rutgers, The State University of New Jersey, Piscataway, NJ, 08855, USA}
\author{E.~Downie}
\affiliation{Department of Physics, The George Washington University, Washington, D.C. 20052, USA}
\author{I.~P.~Fernando}
\altaffiliation{Current address: Department of Physics, University of Virginia, Charlottesville, VA, 22904, USA}
\affiliation{Physics Department, Hampton University, Hampton , VA, 23668, USA}
\author{A.~Flannery}
\affiliation{Department of Physics and Astronomy, University of South Carolina, Columbia, SC, 29208, USA}
\author{T.~Gautam}
\affiliation{Physics Department, Hampton University, Hampton , VA, 23668, USA}
\author{D.~Ghosal}
\affiliation{Department of Physics, University of Basel, 4056 Basel, Switzerland}
\author{R.~Gilman}
\affiliation{Department of Physics and Astronomy, Rutgers, The State University of New Jersey, Piscataway, NJ, 08855, USA}
\author{A.~Golossanov}
\affiliation{Department of Physics, The George Washington University, Washington, D.C. 20052, USA}
\author{B.~F.~Halter}
\affiliation{Department of Physics, The George Washington University, Washington, D.C. 20052, USA}
\author{J.~Hirschman}
\affiliation{Department of Physics, The George Washington University, Washington, D.C. 20052, USA}
\author{Y.~Ilieva}
\affiliation{Department of Physics and Astronomy, University of South Carolina, Columbia, SC, 29208, USA}
\author{M.~Kim}
\affiliation{Randall Laboratory of Physics, University of Michigan, Ann Arbor, MI 48109, USA}
\author{M.~Kohl}
\affiliation{Physics Department, Hampton University, Hampton , VA, 23668, USA}
\author{B.~Krusche}
\affiliation{Department of Physics, University of Basel, 4056 Basel, Switzerland}
\author{I.~Lavrukhin}
\affiliation{Randall Laboratory of Physics, University of Michigan, Ann Arbor, MI 48109, USA}
\altaffiliation{Department of Physics, The George Washington University, Washington, D.C. 20052, USA}
\author{L.~Li}
\affiliation{Department of Physics and Astronomy, University of South Carolina, Columbia, SC, 29208, USA}
\author{B.~Liang-Gilman}
\affiliation{Department of Physics and Astronomy, Rutgers, The State University of New Jersey, Piscataway, NJ, 08855, USA}
\author{A.~Liyanage}
\affiliation{Physics Department, Hampton University, Hampton , VA, 23668, USA}
\author{W.~Lorenzon}
\affiliation{Randall Laboratory of Physics, University of Michigan, Ann Arbor, MI 48109, USA}
\author{P.~Mohanmurthy}
\altaffiliation{Current address: Department of Physics, University of Chicago, Chicago, IL, 60637, USA}
\affiliation{Laboratory for Nuclear Science, Massachusetts Institute of Technology, Cambridge, MA, 02139, USA}
\author{R.~Mokal}
\affiliation{Department of Physics and Astronomy, Rutgers, The State University of New Jersey, Piscataway, NJ, 08855, USA}
\author{P.~Moran}
\altaffiliation{Current address: Laboratory for Nuclear Science, Massachusetts Institute of Technology, Cambridge, MA, 02139, USA}
\affiliation{Department of Physics, Temple University, Philadelphia, PA, 19122, USA}
\author{S.~J.~Nazeer}
\affiliation{Physics Department, Hampton University, Hampton , VA, 23668, USA}
\author{P.~Or}
\affiliation{Racah Institute of Physics, The Hebrew University of Jerusalem, Jerusalem, Israel }
\author{T.~Patel}
\affiliation{Physics Department, Hampton University, Hampton , VA, 23668, USA}
\author{E.~Piasetzky}
\affiliation{School of Physics and Astronomy, Tel Aviv University, Tel Aviv, 69978, Israel}
\author{T.~Rauber}
\affiliation{Paul Scherrer Institute, Villigen, CH-5232, Switzerland}
\author{R.~S.~Raymond}
\affiliation{Randall Laboratory of Physics, University of Michigan, Ann Arbor, MI 48109, USA}
\author{D.~Reggiani}
\affiliation{Paul Scherrer Institute, Villigen, CH-5232, Switzerland}
\author{H.~Reid}
\affiliation{Randall Laboratory of Physics, University of Michigan, Ann Arbor, MI 48109, USA}
\author{G.~Ron}
\affiliation{Racah Institute of Physics, The Hebrew University of Jerusalem, Jerusalem, Israel}
\author{E.~Rooney}
\affiliation{Department of Physics, The George Washington University, Washington, D.C. 20052, USA}
\author{T.~Rostomyan}
\affiliation{Paul Scherrer Institute, Villigen, CH-5232, Switzerland}
\affiliation{Department of Physics and Astronomy, Rutgers, The State University of New Jersey, Piscataway, NJ, 08855, USA}
\author{M.~Schwarz}
\affiliation{Paul Scherrer Institute, Villigen, CH-5232, Switzerland}
\author{A.~Sneath}
\affiliation{Department of Physics, Duquesne University, Pittsburgh, PA, 15282, USA}
\author{P.~G.~Solazzo}
\affiliation{Department of Physics, The George Washington University, Washington, D.C. 20052, USA}
\author{N.~Sparveris}
\affiliation{Department of Physics, Temple University, Philadelphia, PA, 19122, USA}
\author{N.~Steinberg}
\affiliation{Randall Laboratory of Physics, University of Michigan, Ann Arbor, MI 48109, USA}
\author{S.~Strauch}
\affiliation{Department of Physics and Astronomy, University of South Carolina, Columbia, SC, 29208, USA}
\author{V.~Sulkosky}
\affiliation{Laboratory for Nuclear Science, Massachusetts Institute of Technology, Cambridge, MA, 02139, USA}
\author{N.~Wuerfel}
\affiliation{Randall Laboratory of Physics, University of Michigan, Ann Arbor, MI 48109, USA}

\begin{abstract}
The MUon Scattering Experiment, MUSE,  at the Paul Scherrer Institute, Switzerland, investigates 
the proton charge radius puzzle, lepton universality, and two-photon exchange, via simultaneous measurements of elastic muon-proton and electron-proton scattering. The experiment uses the PiM1 secondary beam channel, which was designed for high precision pion scattering measurements.  
We review the properties of the beam line established for pions.
We discuss the production processes that generate the electron and muon beams, and the simulations of these processes.
Simulations of the $\pi$/$\mu$/$e$ beams through the channel using TURTLE and G4beamline are compared.
The G4beamline simulation is then compared to several experimental measurements of the channel, including the momentum dispersion at the IFP and target, the shape of the beam spot at the target, and timing measurements that allow the beam momenta to be determined.
We conclude that the PiM1 channel can be used for high precision $\pi$, $\mu$, and $e$ scattering.
\end{abstract}
\maketitle







\section{Introduction}
\label{intro}
\input{intro}

\section{PiM1 Beam Line}
\label{sec:beamline}
\input{channel}

\section{Source Size}
\input{sources}

\section{Beam Line Simulations}
\subsection{Turtle Simulations}
\input{turtlesim}

\subsection{G4beamline Simulations}
\input{g4beamlinesim}

\section{Comparison of Beam Line Simulations and Measurements}
\input{comp}


\section{Conclusion}
\input{Conclusion}

\section{Acknowledgements}
We acknowledge the Paul Scherrer Institute for its hospitality and support.

This work was supported by the US National Science Foundation (NSF) grants 1436680, 1505934, 1614456, 1614773, 1614850, 1614938, 1649873, 1649909, 1807338, 1812382, 1812402, 1913653, 2012114, and 2012940 by the United States–Israel Binational Science Foundation (BSF) grant 2012032 and 2017673, by the NSF-BSF grant 2017630, by the U.S. Department of Energy (DOE) with contract no. DE-AC02-06CH11357, DE-SC0012589, DE-SC0016577, DE-SC0012485, DE-SC0016583, and DE-FG02-94ER40818, by PSI, Villigen, Switzerland, by Schweizerischer Nationalfonds (SNF) 200020-156983, 132799, 121781, 117601, by the Azrielei Foundation, by the Swiss State Secretariat for Education, Research and Innovation (SERI), Switzerland grant FCS 2015.0594, and by Sigma Xi, USA grants G2017100190747806 and G2019100190747806.

\clearpage

\appendix
\section{Momentum Requirements}
\label{sec:mom_req}
\input{Momentum_requirements}
\section{M Target}
\input{M_Target.tex}




\clearpage

 \bibliography{pim1_beamline.bib}





\end{document}

%% file: intro.tex
An extremely precise measurement of the proton's charge radius using muonic hydrogen found the radius to be $0.84184(67)$ fm \citep{Pohl}.  This measurement was significantly smaller, by $7$ standard deviations, than the accepted charge radius at the time of $0.8775(51)$ fm \cite{CODATA:2010}, which was based on hydrogen spectroscopy and electron scattering measurements.
This discrepancy, known as the Proton Radius Puzzle, was confirmed by contemporaneous electron scattering measurements \citep{Jan,Zhan} and by a subsequent muonic hydrogen measurement \citep{Antognini}. 
This situation led to significant theoretical work on possible explanations along with several new experiments designed to improve our knowledge of the proton's electromagnetic structure \citep{Beyer79,PhysRevLett.120.183001,Grinin1061,ISR,Bezginov1007,PRAD}.

The MUon Scattering Experiment (MUSE) \citep{MUSE} at the PiM1 channel of the Paul Scherrer Institute (PSI) was developed to address the proton radius puzzle through a simultaneous measurement of elastic $ep$ and $\mu p$ scattering, from which the charge radius can be extracted.
The PiM1 channel pion beam has well-studied properties, but, as a secondary beam, has a much larger emittance and momentum bite than modern primary electron beams. This poses a challenge for high precision measurements \citep{Balsiger,Albanese,Hajdas}.
While the properties of electrons and muons from PiM1 might be expected to be similar, as it is a magnetic beam line, these properties have not been well established.
The different production mechanisms for these particles might lead to appreciable differences in the beams from the channel.

In this report, we discuss the PiM1 channel and the known properties of its pion beam.
We present results from the studies undertaken to ascertain the electron and muon beam properties, including source and channel simulations, beam trajectory measurements from our Gas-Electron Multiplier (GEM) planes \cite{Liyanage:2018wup}, and 
time and momentum measurements from our Beam Hodoscope (BH) scintillator paddles \cite{BH}. 
We find that, aside from their fluxes, the pion, muon, and electron beams all have very similar properties.

%% file: channel.tex
\begin{figure*}[htb!]
\includegraphics[width=0.7\linewidth, angle=76]{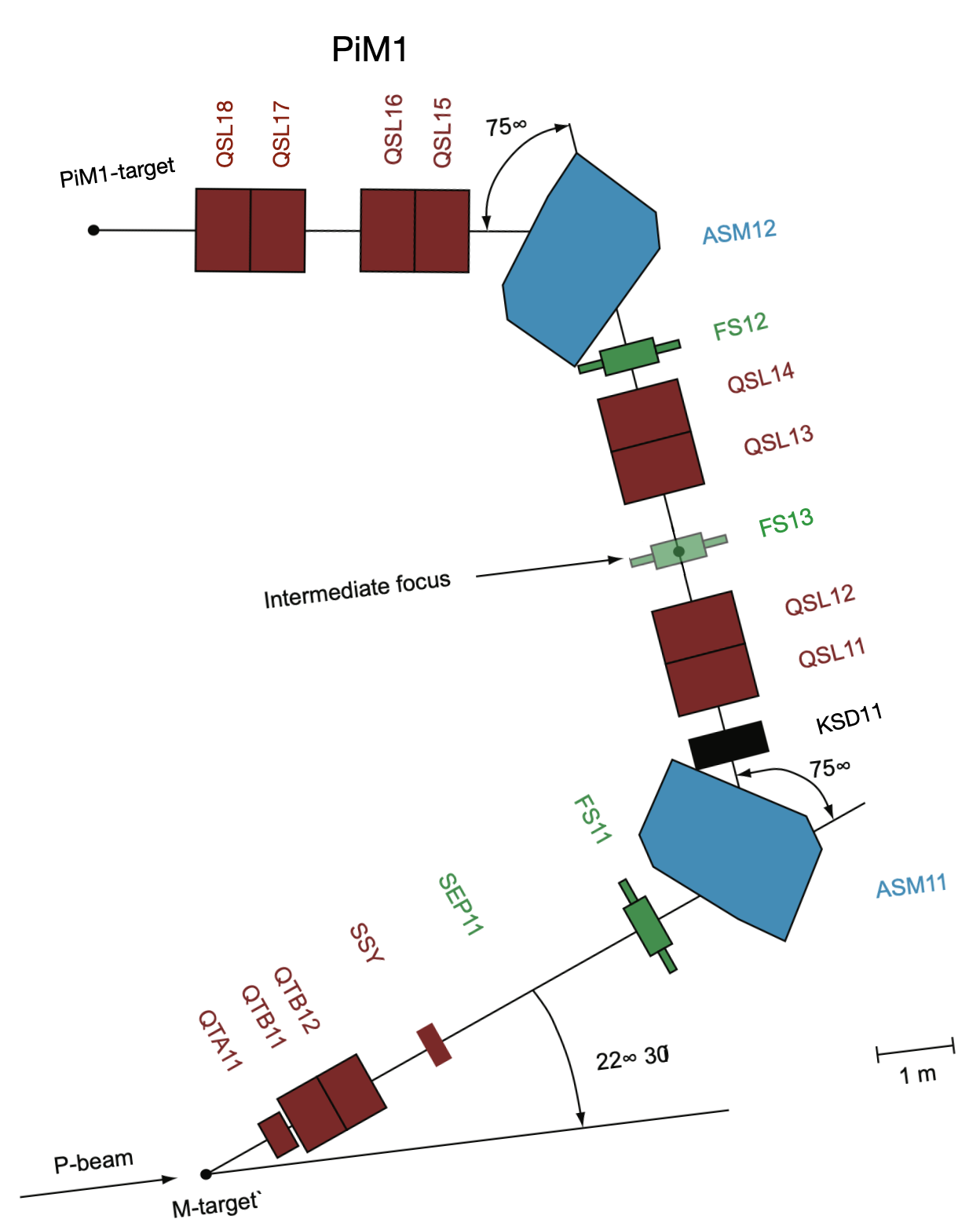}
\caption{\label{fig:MagnetDrawing_PiM1}Drawing of the PiM1 channel. 
ASM11-12, in blue, are dipole bending magnets. 
The first focusing quadrupole triplet, QTA11, QTB11, and QTB12, in red, increases the beam flux.
The quadrupole doublets QSL11-12 and QSL13-14, in red, act in concert with the shaped dipole fields to create a momentum dispersion at the intermediate focus and and then undo it.
The quadrupole doublets QSL15-16 and QSL17-18, in red, focus the beam on the scattering target in PiM1.
There is one vertical steering magnet labelled SSY, in red.
SEP11 indicates the position of an electromagnetic separator, which had formerly been part of the beam line.
Two sets of jaws, FS11 and 12, in green, can be used to limit the flux through the channel. The FS13 jaw, in green, is the collimator placed at the IFP. 
KSD11, in black, is a beam plug, interlocked to allow safe access to PiM1.}
\end{figure*}

The beam line elements of the channel are shown in Fig.~\ref{fig:MagnetDrawing_PiM1}.
The PiM1 channel starts at the M production target. It consists of focusing quadrupoles and two dipoles. 
The proton beam is generated by the High Intensity Proton Accelerator, which produces 590 MeV kinetic energy protons, with a bunch frequency of 50.6 MHz \cite{Seidel:2010zz}. 
The proton beam has a Gaussian distribution both in time and space, with a longitudinal width of $\sigma_L \approx 58$ mm at the M target, corresponding to $\sigma_t \approx 250$ ps width in time \cite{PiM1:privcom}.

The channel, which starts at an angle of approximately 22$^{\circ}$ from the primary proton beam incident upon the M production target, bends particles in a horizontal plane.
The beam height is 1.5 m above floor level.
One vertical bending magnet allows corrections for small vertical offsets in the beam line.
The resulting properties \citep{PiM1} are listed in Table \ref{table:ChannelProp} for a pion source at the production target. 
All channel magnets can be set to either polarity, allowing beams of either charge.
The beam line has two sets of movable apertures, or ``jaws," located at FS11 and FS12 in Fig.~\ref{fig:MagnetDrawing_PiM1}.  Each jaw set consists of both horizontal and vertical jaw pairs;
each individual jaw can be independently set from approximately the beam axis out to beyond the beam envelope. 

An important feature of the beam line is an intermediate focal plane (IFP), where the beam is momentum dispersed, allowing the momentum to be better determined. 
This feature allowed PiM1 to be used for pion scattering studies requiring high momentum resolution.
The full $\pm$1.5 \% momentum bite is spread out over $\pm$10.5 cm. 

\begin{table*}[htb!]
\caption[PiM1 channel properties]{
PiM1 channel properties determined previously for pions \cite{PiM1}. 
\label{table:ChannelProp}}
\begin{tabular}{ c@{\hskip 0.5in}c } 
 \hline
 \hline
{\bf Property} & {\bf Value}  \\ 
 \hline
Total Path Length & $23.12$ m \\
 Momentum Range & 100$-$500 MeV/$c$ \\
 Solid Angle & $6$ msr \\
 Momentum Acceptance (FWHM) & $\pm 1.5\%$ \\
 Momentum Resolution & $0.1\%$ \\
 Dispersion at the IFP & $7$ cm/\% \\
 Spot Size on Target (FWHM) & $15$ mm horizontal by $10$ mm vertical \\
 Angular Divergence on Target (FWHM) & $35$ mrad horizontal by $75$ mrad vertical \\
 \hline
 \hline
\end{tabular}
\end{table*}

\begin{figure*}[htb!]
\hspace{-6em}\includegraphics[width=\linewidth]{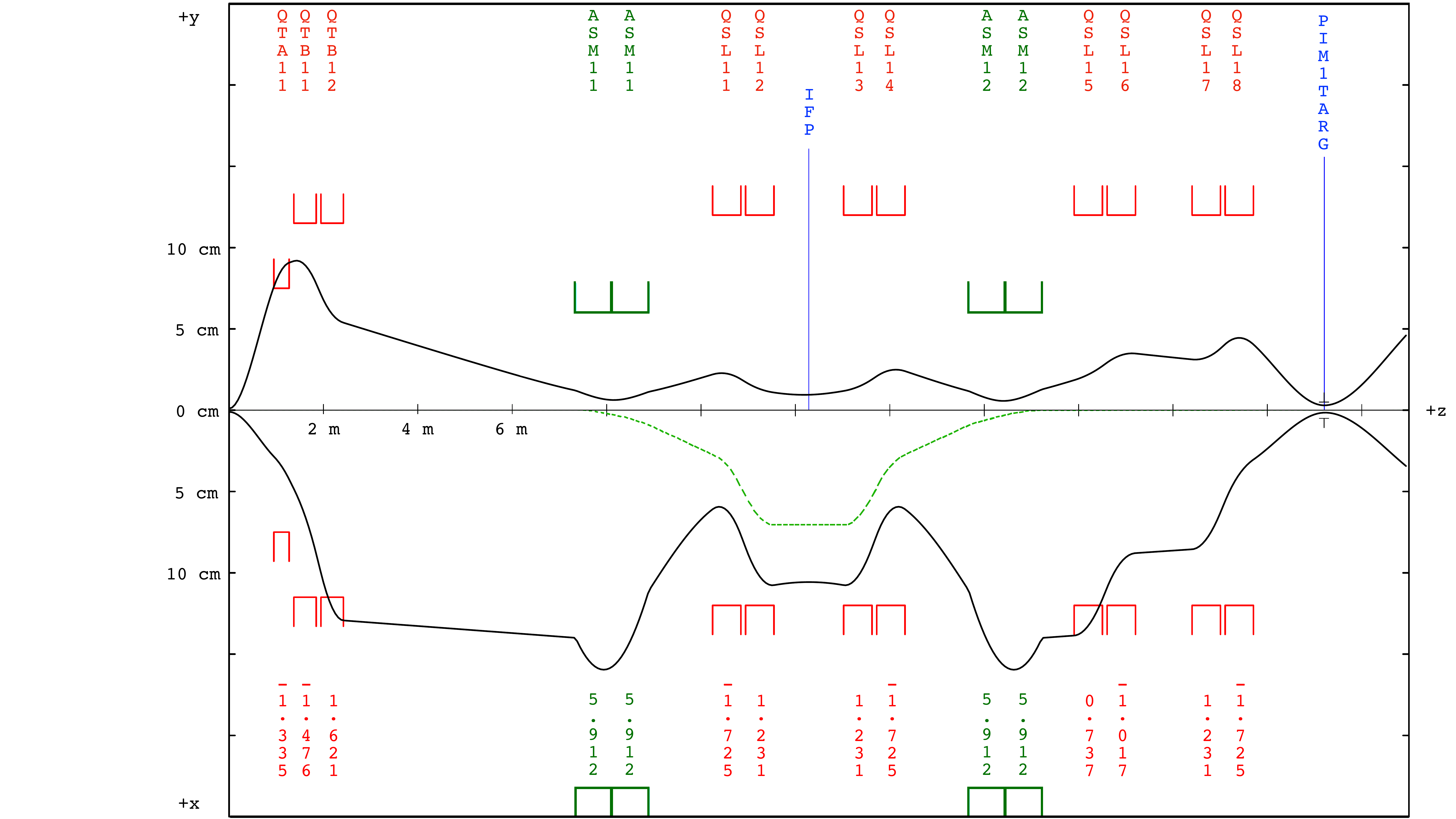}
\caption[Transport calculation of the PiM1 channel]{
The TRANSPORT calculation of the PiM1 channel. 
The red open rectangles represent the quadrupole positions and apertures.
The dark green open rectangles represent the dipole positions and apertures. 
The numerical values at the bottom show the magnetic field polarities and magnitudes (kG).
The nominal position of the MUSE scattering target in PiM1 is labelled \enquote{PIM1TARG}. 
\enquote{DR5} represents the position of the FS11 jaws. 
The two black lines show the 2$\sigma$ beam envelope in the vertical $y$ direction in the top half of the figure, and in the horizontal $x$ direction in the bottom half of the figure.
The green line visible near the IFP shows the momentum dispersion of the beam (in cm/\%, rather than cm).
Vertical tick marks are positioned every 5 cm. Horizontal tick marks are positioned every 2 m.}
\label{fig:Transport_PiM1}
\end{figure*}

A more detailed picture of the beam as it progresses through the channel can be seen with a TRANSPORT calculation \cite{Transport}, shown in Fig.~\ref{fig:Transport_PiM1}. 
Of particular note here is the IFP region.
The symmetry of the quadrupoles and dipoles around the IFP generates and removes the momentum dispersion in the beam.
The IFP itself is within an air gap about 60 cm long, in the middle of the beam line. This facilitates the installation of detectors and jaws, although the airgap is not included in the standard TRANSPORT calculation.
To be able to limit the channel momentum bite, and to enable more detailed beam studies, an adjustable copper collimator, FS13, was installed at the IFP.
Unlike the FS11 and FS12 jaws, the FS13 opening can be centered at any point across the dispersed beam at the IFP, as the copper blocks can cross the beam axis.

\begin{table*}[htb!]
\caption[PiM1 Matrix Elements from Production Target]{
First-order PiM1 beam transport matrix elements from the production target to the IFP and to the PiM1 scattering target, calculated with TURTLE. 
}
\label{tab:1storderme}
\begin{tabular}{ cccccc } 
 \hline
 \hline
             & $x_{prod}$  & $x'_{prod}$   & $y_{prod}$ & $y'_{prod}$ & $\delta$ \\ 
 \hline 
 $x_{\text{IFP}}$   &  $-$1.221      & $-$0.036 cm/mrad & $-$0.079       & 0            & 7.07 cm/\% \\
 $x'_{IFP}$  &  $-$3.37 mrad/cm & $-$0.919       & $-$0.005 mrad/cm & 0            & 1.276 mr/\% \\
 $y_{IFP}$   &  0          & 0             & 9.65        & $-$0.009 cm/mrad & 0 \\
 $y'_{IFP}$  &  0          & 0             & 52.04 mrad/cm  & 0.052        & 0 \\
 \hline
 $x_{targ}$  &  1.47       &  0   & 0.325        & 0            & $-$0.137 cm/\% \\
 $x'_{targ}$ &  7.31 mrad/cm & 0.678  & 0.967 mrad/cm   &0       & $-$2.01 mrad/\% \\
 
 $y_{targ}$  &  0          & 0             & $-$0.319       & $-$0.004 cm/mrad & 0 \\
 $y'_{targ}$ &  0          & 0             & 273.8 mrad/cm  & $-$0.091       & 0 \\
 \hline
 \hline
\end{tabular}
\end{table*}

The first-order matrix elements for the PiM1 channel were calculated using TURTLE \cite{Turtle,turtlemanual,fringeFields}, which uses as an input the same description of the beam line geometry and magnetic elements as TRANSPORT.
The calculation was done using a non-divergent beam from a point-like source, {\em i.e.}, with $(x,x',y,y',\delta)$ all at constant values at the M production target. Here $x$ and $x'$ are the horizontal position and angle respectively, with beam right being the positive direction, $y$ and $y'$ are the vertical position and angle, with vertically up being positive, and $\delta$ represents the relative momentum of the channel as a percent of the central momentum. Then one source parameter at a time was varied while all others were kept at zero, to ensure the resulting offsets at the IFP and PiM1 target used in determining the matrix elements were still in the linear regime.

Table~\ref{tab:1storderme} shows the resulting first-order matrix elements from the source to the IFP and to the PiM1 target.
The $\left< x_\mathrm{IFP} \, | \, \delta \right>$ matrix element can be seen to be 7 cm/\%.
The $x$, $x'$ and $\delta$ coordinates are largely decoupled from the $y$, $y'$ coordinates.
Pions of the same momentum are spread out at the IFP by $\pm$0.72 cm. This mainly arises from the $\pm$20 mr acceptance of the channel for pions, along with the matrix element of $-0.036$ cm/mr from the production target to the IFP.
Comparing to the 7 cm/\% dispersion at the IFP, we see a $\pm$0.1 \% range of momenta at any $x_{IFP}$ position.

\begin{table*}[htb!]
\caption[PiM1 Matrix Elements from IFP]{
First-order PiM1 beam transport matrix elements from the IFP to the PiM1 scattering target, calculated with TURTLE. 
}
\label{tab:1stordermeifp}
\begin{tabular}{ cccccc } 
 \hline
 \hline
             & $x_\mathrm{IFP}$  & $x'_\mathrm{IFP}$   & $y_\mathrm{IFP}$ & $y'_\mathrm{IFP}$ & $\delta$ \\*[3pt]
 \hline 
 $x_\mathrm{targ}$   &  $-$1.35      & $-$0.05 cm/mrad & $-$0.001       & 0            & 9.52 cm/\% \\
 $x'_\mathrm{targ}$  & $-$4.44 mrad/cm & $-$0.573       & $-$0.003 mrad/cm & 0            & 30.3 mrad/\% \\
 $y_{\mathrm{targ}}$   &  0          & 0             & 0.17        & $-$0.04 cm/mrad & 0 \\
 $y'_{\mathrm{targ}}$  &  0          & 0             & 19.01 mrad/cm  & 0.04        & 0 \\
 \hline
 \hline
\end{tabular}
\end{table*}

The air gap at the IFP leads to a small energy loss and a few mrad of multiple scattering, which will affect the beam properties after the IFP.
To the degree to which there is simply a slight momentum shift, it is, in normal operation, entirely compensated for through a slight adjustment of the ASM dipole magnet settings.
To judge the actual effects of material in the beam, it is important to consider the matrix elements from the IFP to the PiM1 target. Table~\ref{tab:1stordermeifp} shows these matrix elements calculated with TURTLE, by removing from the input all channel elements before the IFP.
Again, the $x$, $x'$ and $\delta$ coordinates are largely decoupled from the $y$, $y'$ coordinates.
The effect of a few mrad of multiple scattering is to enlarge the beam spot at the PiM1 target by a few mm.
The important point to consider is the dispersion at the PiM1 target from the IFP, $\left< x_\mathrm{targ} \, | \, \delta \right>$ = 9.52 cm/\%.
Of course, the position and momentum of particles at the IFP from the source are correlated so that the dispersion at the target is small, $\left< x_{targ} \, | \, \delta \right>$ = $-0.137$ cm/\%.
But the air gap at the IFP induces a differential momentum loss of the particle species, shifting the beam spots in PiM1. The shift between the beam spots of the particle species in PiM1 is a sensitive measure of any momentum differences between the particles.
The proton energy loss is so large that the channel cannot simultaneously transport protons to PiM1 along with the electrons, muons and pions with the MUSE beam tunes\footnote{It is possible to tune the beam in such a way that protons can be transported, in which case all electrons, muons, and pions are lost.}.

%% file: sources.tex
Modeling the distributions of pions, muons and electrons at the PiM1 target not only requires knowledge of the  PiM1 channel, but also knowledge of the proton beam, the M production target geometry, and the physics processes that generate the various particle types.
It is inefficient to simulate the secondary beam source sizes starting from primary beam protons. We have adopted the approach of modeling the secondary beam sources with separate source simulations that use an approximate PiM1 channel acceptance.
First, we discuss the channel acceptance, then we turn to the various particle sources.

\subsection{Channel Acceptance}
\input{acceptance}

\subsection{Pion Source}
\label{sec:PionSource}

The M production target is a rotating graphite wheel positioned such that protons transit a 2-mm wide flange, approximately aligned to the PiM1 beam line, at an angle of 22$^{\circ}$ from the primary proton beam line. 
The size of the proton beam and the flange it crosses leads to a simple approximation for the pion (point-like) source: it is distributed across a 2 mm $\times$ 2 mm full-width spot in $x_{prod}$ $\times$ $y_{prod}$.
The approximately $\pm$3 mm variation in $z$ across the beam spot is small compared to the 96 cm distance to the first quadrupole, QTA11, and is ignored.
The beam is assumed to be uniform in momentum over the $\pm$1.5 \% momentum bite of the channel.

The pions of interest to MUSE are generated with momenta between 115 and 210 MeV/$c$, corresponding to kinetic energies between approximately $40$ and $112$ MeV. 
A larger spatial distribution needs to be considered for generation of muons in the same momentum range.
The production of pions by the 590 MeV proton beam striking the target M was studied in Ref. \citep{PiProd}.
The few percent variation in flux across the channel acceptance does not significantly effect the beam spot or timing measurements. 
Our pion simulations use this simple, uncorrelated source.
Due to this simple approximation, distributions that have a significant  correlation with momentum will have appropriate limits but not the right shape.

\subsection{Muon Source Simulations}
\label{sec:MuonSource}

\begin{figure*}[htb!]
\includegraphics[width=\linewidth]{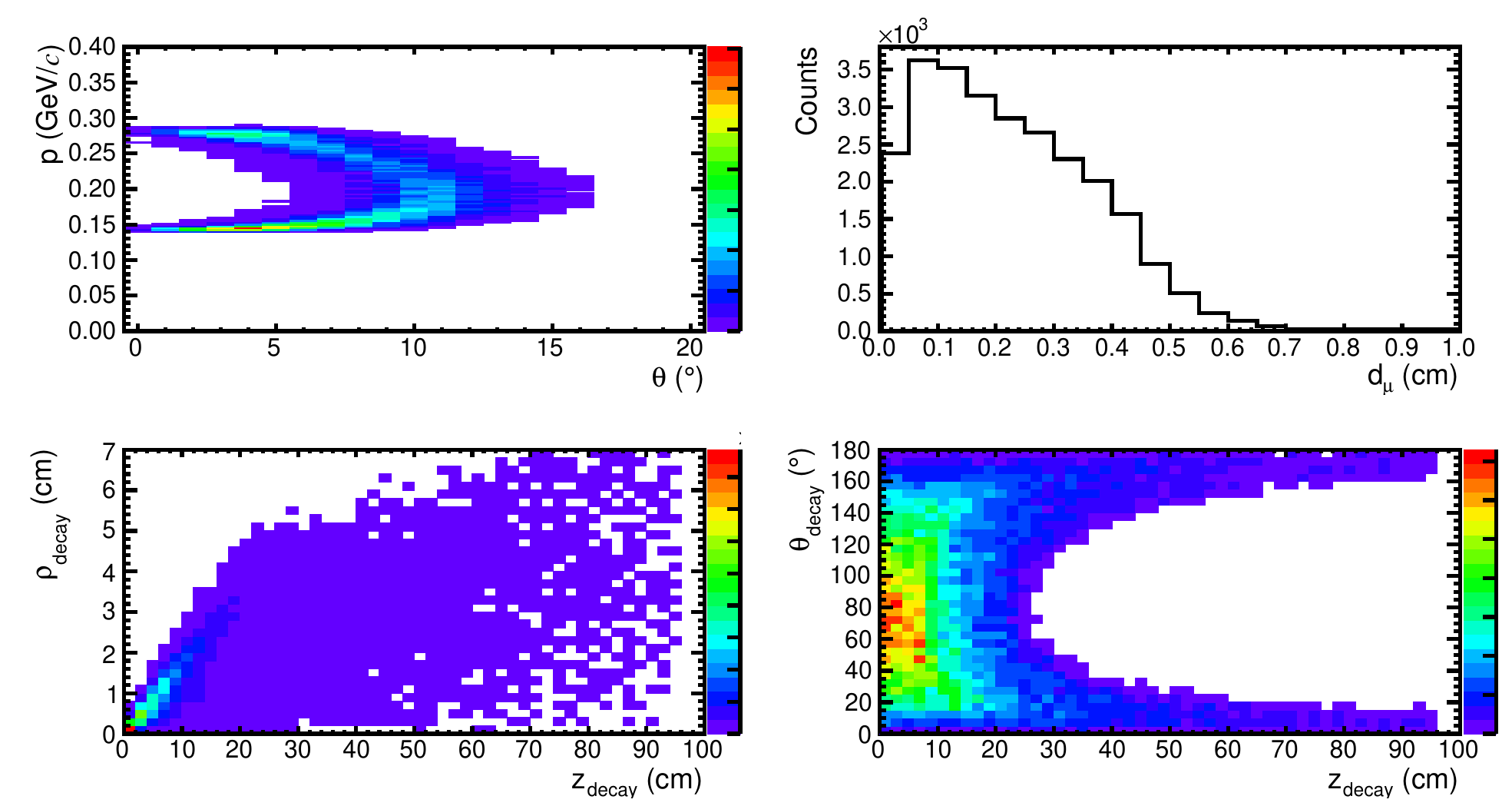}
\caption{\label{fig:210muonsfromroot}Output of a ROOT simulation of the muon source size, for 160 MeV/$c$ channel setting.
\textit{\textbf{Upper left}}: Momentum in GeV/$c$ vs angle in degrees of $\pi$'s that generate muons that pass the first quad aperture cuts.
\textit{\textbf{Upper right}}: Extra distance traversed by the combined decay pions and generated muons compared to a straight trajectory from the pion source point to point of entrance at the front face of QTA11.
\textit{\textbf{Lower left}}: Distribution of $\pi$ decay point transverse distance $\rho$ vs. z for pions that decay to muons and traverse the beam line to the PiM1 target.
\textit{\textbf{Lower right}}: Distribution of $\pi$ decay $\theta$ vs. z for pions that decay to muons and traverse the beam line to the PiM1 target.
The 2D histograms are shown on a linear scale.}
\end{figure*}

Charged muons are produced via the $\pi \to \mu\nu$ decay.
The physics input to the decay is straightforward as the decay is isotropic in the pion rest frame and the pion life time is well known.
For the production of muons that go into PiM1 in the momentum range of interest, we are interested in pions in a somewhat more extended momentum range that decay within approximately 15$^{\circ}$ (25$^{\circ}$) of the PiM1 channel for the highest (lowest) MUSE momenta.
In this case, it is sufficient to assume a pion spectrum that is uniform in kinetic energy and angle. 

The muon simulation used the pion source discussed in Sec.~\ref{sec:PionSource}
and the cuts discussed in Sec.~\ref{Sec:ChannelAcceptance}.
The muons are generated over a wider range than the channel accepts so that the results are determined by the acceptance of the channel, not by using a limited muon source.
The overall efficiency of producing muons into PiM1 from the source pion distribution described is $\mathcal{O}({10^{-5}})$.
Some results from a simulation of 2$\times$10$^9$ pions are shown in Fig.~\ref{fig:210muonsfromroot}, for a channel central momentum of $160$ MeV/$c$. 
The top left panel shows the relation between pion momentum and angle for pions that decay to muons that make it through QTA11 and the correlation cuts.
The top right panel shows that the combined flight path of the generated pion and decay muons is about 2 mm longer than a direct flight path into the channel -- the muons travel little extra distance compared to the pions.
The bottom left panel shows how far from the channel central axis the muons were created, as a function of the distance along the $z$-axis.
The bottom right panel shows the decay angle in the pion rest frame as a function of the distance along the axis.
These plots confirm that most of the muons passing channel acceptance cuts come from pions with trajectories at angles near 10$^{\circ}$ from the channel axis, that decay in their first several cm of flight and within 1 cm of the channel axis.

The simulations were confirmed with a more statistically limited Geant4 simulation, which indicated that the effects of the Target M graphite wheel on the muon distributions were small.
As a result, ROOT was used to generate muon rays to feed into the beam line simulations.
Approximately $10^{11}$ pions were generated, leading to nearly $1.4\times 10^6$ muon rays that were likely to be transported to the PiM1 target by the channel.
Note that, in practice, the probability that the preselected likely-to-succeed rays were successfully transported was typically several percent, depending on the details of the PiM1 channel simulation.

\subsection{Electron/Positron Source Simulations}
\label{sec:EpEmSource}

\begin{figure*}[htb!]
\includegraphics[width=\linewidth]{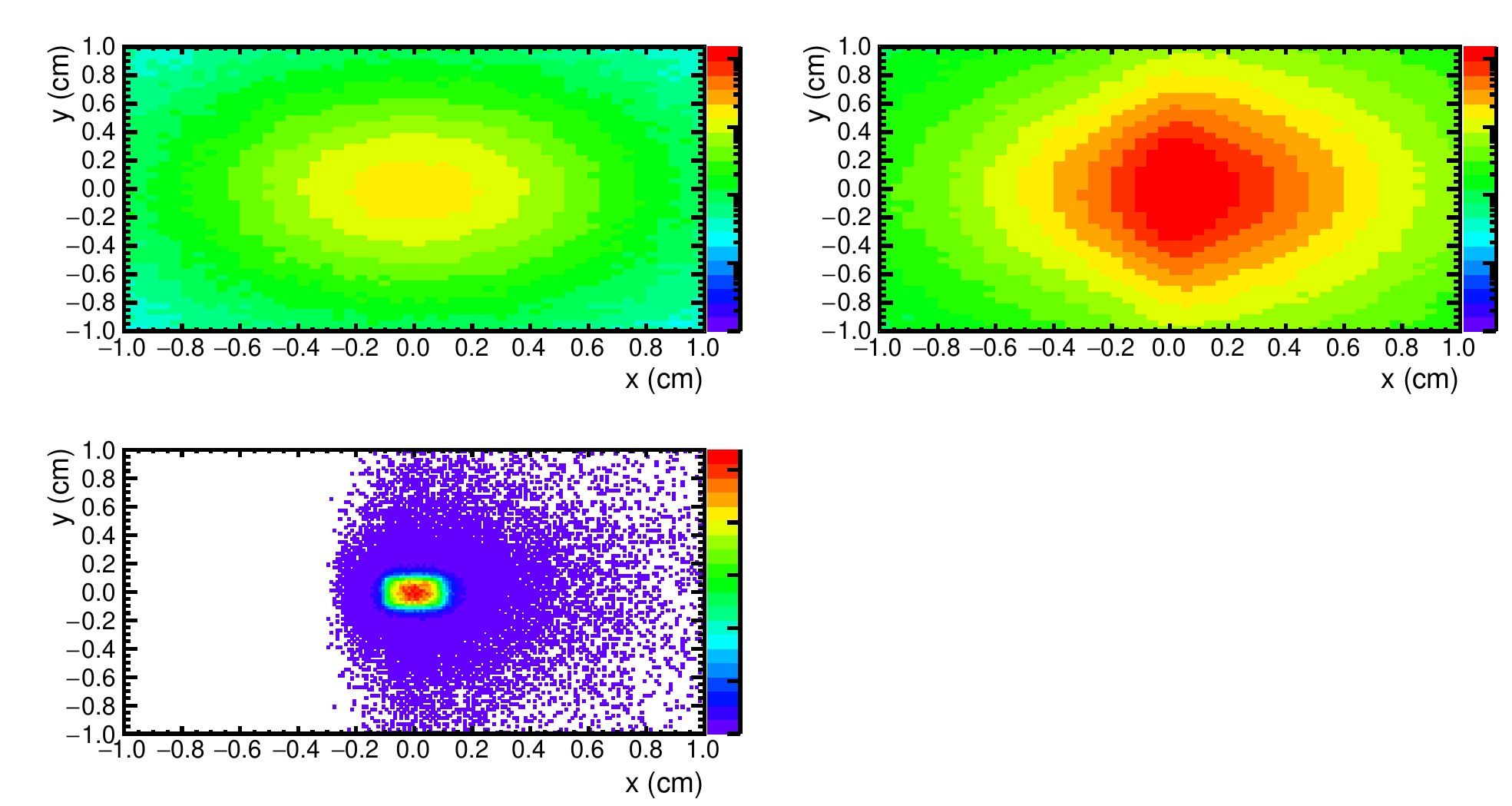}
\caption{\label{fig:epdistsgraphite}Geant4 simulated electron/positron distributions on a test plane immediately after the M target graphite wheel, at $z = 2.55$ cm.
\textbf{\textit{Top Left:}} The distribution when the simulation is performed without the graphite wheel, so that only Dalitz decays are shown. 
\textbf{\textit{Top Right:}} The full simulated distribution including both $\gamma$ conversion in the graphite wheel and Dalitz decays.
Both left and right are shown on the same logarithmic scale.
\textbf{\textit{Bottom Left:}} Electron/positron source distributions for particles satisfying QTA11 cuts and correlations. The color grading is shown on a linear scale.
}
\end{figure*}

The production of $e^{\pm}$ arises primarily from the decay of $\pi^0$ produced in the $pC \rightarrow \pi^0 X$ reaction.
The branching ratio of the $\pi^0 \to \gamma\gamma $ decay is approximately 98.8 \%.
The produced $\gamma$'s can then interact with the target M to either pair produce $e^+e^-$, or Compton scatter and knock out electrons.
Dalitz decay, $\pi^0 \to e^+e^-\gamma$ has a branching ratio of aproximately 1.2 \%.
The short $\pi^0$ lifetime, with $c\tau$ = 25 nm, results in the decays happening in the beam spot at negligible distance from the production point.
As a result, the Dalitz decay component of the source is point-like, the same as the pion source.

The $e^{\pm}$ source simulation was implemented in Geant4, including the physics described above and the geometry of the target M graphite wheel.
We did not attempt to simulate the $p C \to \pi^0 X$ reaction, due to the significant computational time required \cite{PiProd}. 
We instead assumed that the cross section for production of $\pi^0$'s is approximately constant with kinetic energy and angle, as we had assumed for the production of $\pi^{\pm}$ in the muon source simulations.
The $e^{\pm}$ rays produced were then subject to the same source constraints as the muon rays of Sec.~\ref{sec:MuonSource}.

The electron distributions generated in Geant4 are seen in Fig.~\ref{fig:epdistsgraphite}, evaluated at $z = 2.55$ cm, directly after the production target.
Without the graphite disk, there is no $\gamma$ conversion so only the Dalitz decay generates $e^{\pm}$. 
This is seen in the top left panel, which shows all generated $e^{\pm}$ when there is no graphite wheel.
The right panel of the figure shows the Geant4 simulation including the graphite wheel, which allows $\gamma$ conversion.
The scale is the same for both top left and top right panels, so the $e^{\pm}$ produced from $\gamma$ conversion are about an order of magnitude greater than those from Dalitz decays.
The left-right asymmetry of the distribution in the upper right panel is due to the graphite disk, which leads to the $\gamma$ conversion occurring in the $+x$ side of the figure. 
The bottom left panel of Fig.~\ref{fig:epdistsgraphite} shows the remaining source distributions immediately after the target M, at $z = 2.55$ cm, for events that pass the QTA11 cuts.
The resulting source distribution has an RMS width of approximately 0.4 mm in both $x$ and $y$ directions.
Overall, even though the Dalitz decays account for only a 1.2\% branching ratio, we find that after the QTA11 cuts these decays account for about 16\% of the $e^{\pm}$ going into the PiM1 channel.

To further understand why the $e^{\pm}$ are not generated over a larger volume of the M production target, we investigated the momentum, angle, and position correlations in the electron sources with no momentum or angle cuts on QTA11.
We find that the $e^{\pm}$ momenta drop rapidly as the photon conversion position moves away from the origin transverse to the $z$ axis.
The kinematic reason for this is that the decay products will have high momenta only if aligned to the decaying particle, which limits decays at large angles and consequently any growth in the transverse size of the source.

One limitation to the $e^{\pm}$ source simulation is that for simplicity the initial $\pi^{0}$ distribution was generated at a point, rather than across the entire flange. 
Since the electron source distribution appears to be so much smaller than the flange width for a point source, we conclude that it is sufficient to use the \textit{same} source size for both electrons and pions. 

\subsection{Source Simulations Summary}

We have concluded that the $e^{\pm}$ and $\pi^{\pm}$ source sizes are small, consistent in size with the proton beam spot and can be considered ``point-like". 
Consistent source sizes are not sufficient to ensure consistent behavior of the particle species in the magnetic channel. 
There are three noteworthy mechanisms that can lead to different behaviors despite consistent source sizes.
First, the production mechanisms generate different momentum-angle correlations, so there will be some difference to the shape of the distributions if these correlations are important. We do not find this to be a significant effect in the measured data.
Second, more importantly, there is material in the beam, in particular in the IFP region, that causes multiple scattering and energy loss.
These effects can generate differences between particle species at the MUSE target even if there were no differences at the IFP.
This was mentioned previously in the discussion of the PiM1 matrix elements in Sec.~\ref{sec:beamline} and experimental results are shown in Sec.~\ref{sec:comp}.
Finally, the pion beam properties are somewhat distorted in measurements if muons from $\pi^{\pm}$ decay in flight along the channel are identified as still being pions.
At MUSE momenta, pions decay at a rate of approximately 10 \%/m, which, depending on momenta and measurement technique, can lead to an appreciable background. 
The muon beam in contrast has a ``large" source, and potential differences that need to be characterized by the full beam line simulations.

%% file: acceptance.tex
\label{Sec:ChannelAcceptance}

\begin{figure*}[htb!]
\includegraphics[width=\linewidth]{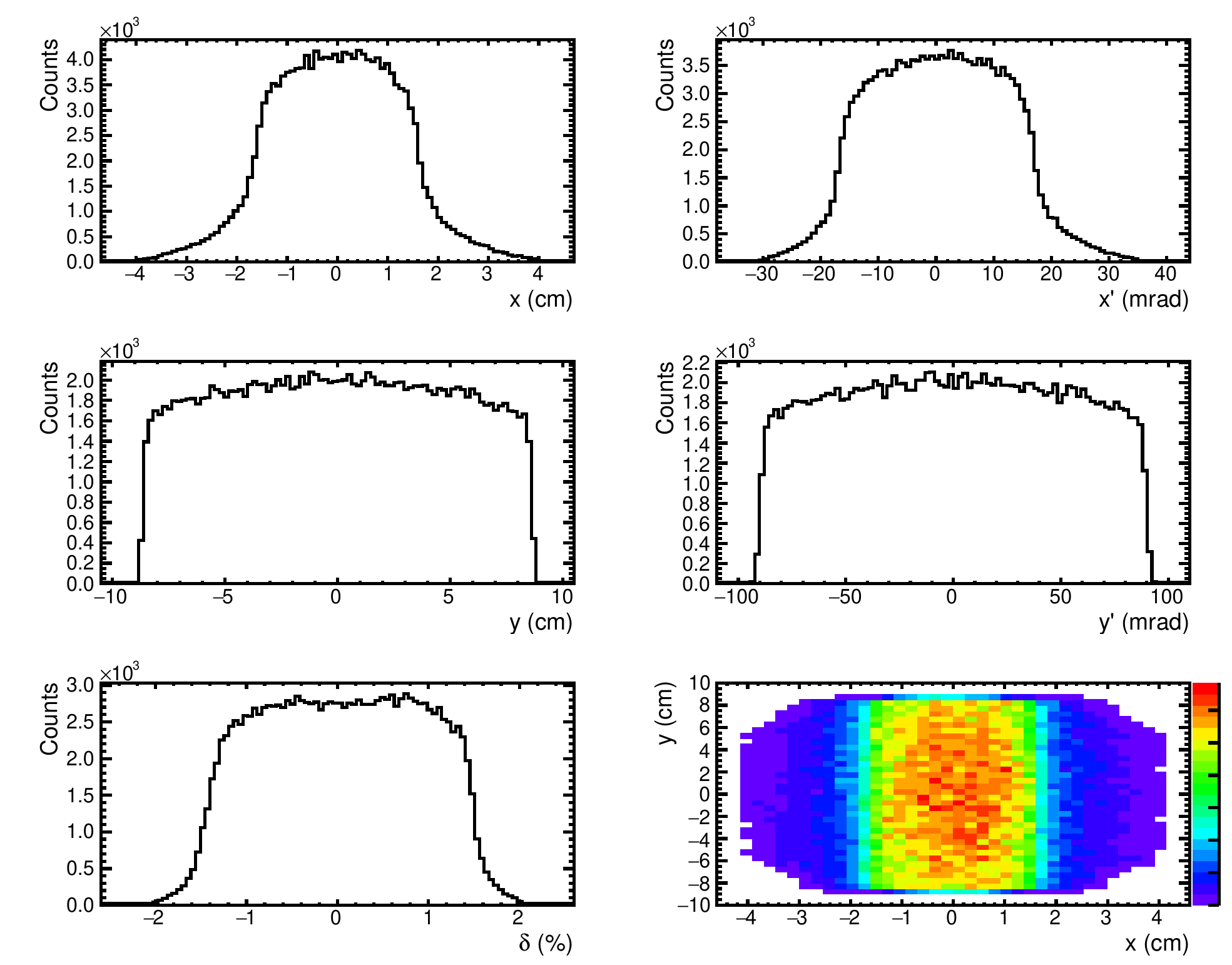}
\caption{Distributions from TURTLE of beam particles at the entrance to the PiM1 channel, for particles transported through the whole PiM1 channel to the GEM chambers. \textit{\textbf{Upper left}}: position distribution in $x$. \textit{\textbf{Upper right}}: $x'$, the angular distribution in $x$. \textit{\textbf{Center left}}: position distribution in $y$. \textit{\textbf{Center right}}: $y'$, the angular distribution in $y$. \textit{\textbf{Lower left}}: momentum distribution $\delta$ in \% from nominal momentum. \textit{\textbf{Lower right}}: position distribution $y$ vs $x$ on a linear scale.}
\label{fig:qtadistcut}
\end{figure*}

\begin{figure*}[htb!]
\includegraphics[width=\linewidth]{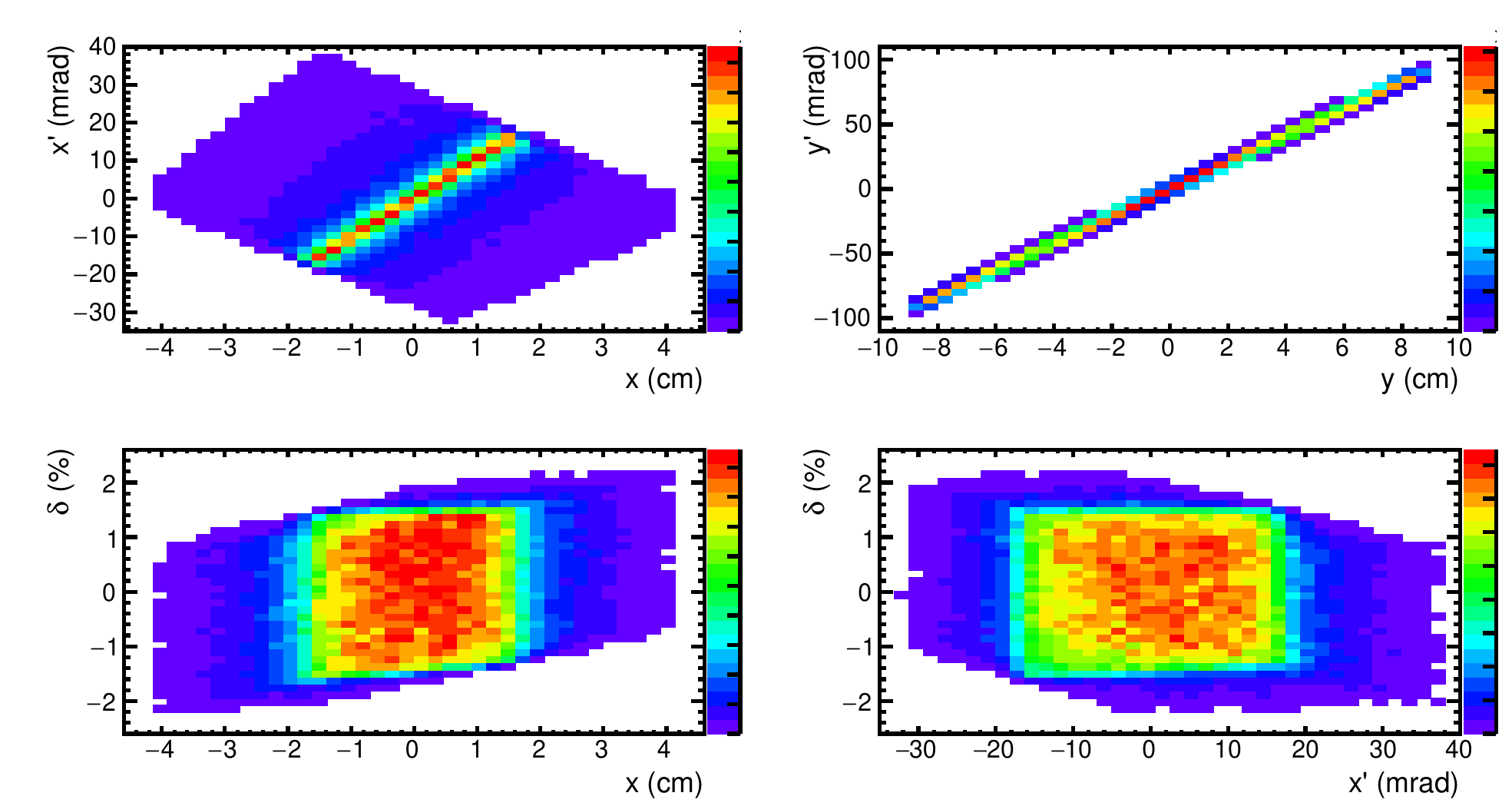}
\caption[QTA1 entrance correlations]{Correlations of particles at the QTA11 entrance that reach the GEM chambers after the PiM1 channel from the TURTLE simulation with a large source. \textit{\textbf{Upper left}}: $x'$ vs $x$. \textit{\textbf{Upper right}}: $y'$ vs $y$. \textit{\textbf{Lower left}}: $\delta$ vs $x$. \textit{\textbf{Lower right}}: $\delta$ vs $x'$. No correlation is observed between relative momentum and either $y$ or $y'$. All panels are shown on a linear scale.}
\label{fig:qtacorrcut}
\end{figure*}

The proton beam crosses the M production target with a spot size of $\pm$1 mm in $x$ and $y$ as seen by the PiM1 channel.
TURTLE simulations with this source size (see Sec.~\ref{sec:turtlesim}), find trajectories accepted for a range of about $\pm$17 mrad in $x'$ $\times$ $\pm$90 mrad in $y'$, corresponding to 6.1 msr.  
The $\pm$1 mm source size, appropriate for pions, is considered to be point-like.
As input to our simulations of the possibly extended sources of muons and electrons, it is important to consider what potential range of particle trajectories might be accepted by the channel.
A TURTLE simulation run with much larger, uncorrelated source rays found acceptance out to 
$\pm$5 cm in $x$, 
$\pm$35 mrad in $x'$, 
$\pm$0.5 cm in $y$, 
$\pm$90 mrad in $y'$, and
$\pm$2 \% in $\delta$.
Note that the source size is not much changed in the $y$, $y'$ directions, but correlations between $x$, $x'$ and $\delta$ allow the source to be significantly larger in the $x$ direction, with twice the angle acceptance and 1/3 larger momentum acceptance.

To implement a larger but still efficient source, we also considered whether additional constraints can be applied based on the beam at the entrance to the first magnet, QTA11.
Figure~\ref{fig:qtadistcut} shows the distribution of beam particles at QTA11 from the large source TURTLE simulation.
All rays shown passed through the channel and reached the PiM1 target region -- more specifically the 10 cm $\times$ 10 cm MUSE GEM tracking chambers 30-cm upstream of the target position that are used in the experiment to measure beam particle trajectories.
QTA11 de-focuses (focuses) in the $x$ ($y$) direction.
As a result, the $\pm4$ cm ($\pm$9 cm) wide acceptance in $x$ ($y$) distribution is narrower (wider) than the 7.5 cm radius of the quadrupole.

A more stringent constraint can be imposed by looking at the correlations between $x$, $x'$ and $\delta$, and between $y$ and $y'$.
Figure~\ref{fig:qtacorrcut} shows these correlations at the entrance to the channel. 
There is a strong $y'$ vs.\ $y$ correlation at the channel entrance, reflecting that particles from outside a several-mm vertical region are not successfully transported through the channel.
There is a similar correlation in $x'$ vs.\ $x$ for a vast majority of the particles, but there remains some acceptance over a much wider phase space.

Cuts based on distributions of particles transported to the PiM1 target from a ``large'' source, shown in Fig.~\ref{fig:qtadistcut}, and the correlations seen in Fig.~\ref{fig:qtacorrcut}, were applied to the source simulations so that sets of particle trajectories more likely to be successfully transported through the PiM1 channel were input to the beam line simulations.

%% file: turtlesim.tex
\label{sec:turtlesim}

The standard TURTLE simulation uses the same input as the TRANSPORT simulation presented in Sec.~\ref{sec:beamline}.
To ensure an accurate description of the beam line, in particular to investigate possible apertures that could affect the tails of the muon distributions, the standard PSI description of PiM1 was reexamined, resulting in several minor changes, including the following:
\begin{itemize}
	\item Minor modifications were made to the placement of elements along the beam line.
	\item Apertures were added to the simulation to simulate the effects of beam pipes and flanges which had not previously been in the description.
	\item The magnetic length of the QSL quadrupoles was reduced from 60 cm to 59 cm to better reflect field maps.
	\item The QSL15-18 fields are tuned for a good focus at the MUSE target. The simulation included fields corresponding to the set currents.
	\item An aperture was added at the IFP to simulate the effect of the new FS13 collimator.
\end{itemize}
The effects of these changes on the simulation were all small.
There are two remaining, known issues with the simulation, for which the resolutions are not clear.

First, in normal operation the SSY steering magnet must be used to center the beam in PiM1 vertically, at the 1.5 m height. 
It is believed that this is needed to counteract a small vertical offset in the first quadrupole triplet of QTA11, QTB11 and QTB12.
These vertical offsets are ignored in the simulation.

Second, the first quadrupole triplet is located in a ferromagnetic, iron enclosure.
This is known to affect their magnetic properties.
The default simulation uses a set of fields that lead to a good description of the beam spot.
A simulation where the fields correspond to the set currents does not. 
Our simulations use the default fields provided by the TURTLE description of the beam line.

Our standard TURTLE simulation does not include materials at the IFP or particle decays.
Particles striking apertures are lost, rather than having multiple scattering and energy loss.
Since PiM1 is a magnetic channel, the TURTLE simulations predict that the electron and pion distributions are the same and are channel momentum independent.
Since the muon simulation uses muon rays generated as described in Sec.~\ref{sec:MuonSource}, the predicted distributions differ from those of electrons and pions.

\subsection{TURTLE Simulations from the IFP to the PiM1 Target.}
\label{section:dispersion}

In Sec.~\ref{sec:beamline} we documented the PiM1 matrix elements, which allow us to study the beam momentum through the particle distributions at the target. However, since this description of the beam line has the production target correlations built-in, we also investigate how a more general and random distribution at the IFP impacts the distribution at the target. This random distribution at the IFP allows for a more thorough understanding of how the momentum spread at each point in the IFP affects the distribution at the target. The generated distributions at the IFP match the size of the point distribution beam at the IFP, with  $\pm$10.5 cm in $x$, $\pm$20 mrad in $x'$, $\pm$2.5 cm in $y$, $\pm$15 mrad in $y'$, and $\pm$1.5 \% in $\delta$, but lack correlations between parameters. We consider the case of the default magnet settings.

The $y$ and $y'$ distributions arising from these simulations are uncorrelated with $\delta$.
This is consistent with the matrix elements shown previously. 
For brevity we omit other plots showing no correlation of $y$ or $y'$ with $\delta$.

\begin{figure*}[htb!]
\centering
\includegraphics[trim=0 0 0 0,clip,width=\linewidth]{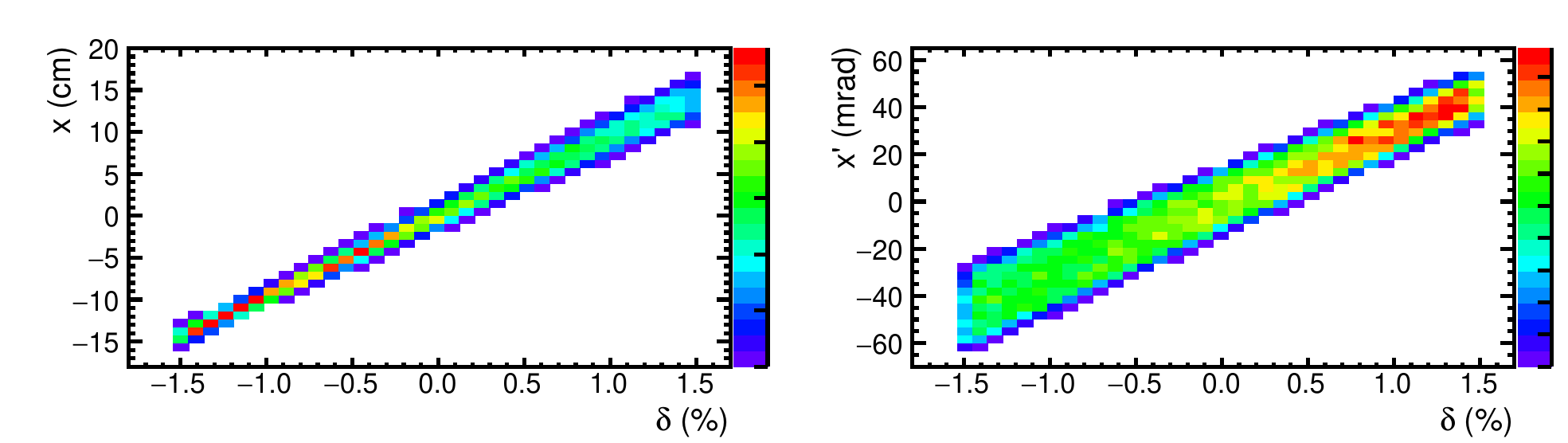}
\caption[Simulations of PiM1 $x$ correlations at 3 IFP momentum bites]{Simulations of position and momentum correlations from TURTLE. \textit{\textbf{Left}}: Dispersion in $x_{targ}$ in cm/\% for particles at $x_{IFP}$ = 0 $\pm$ 0.5 cm.
\textit{\textbf{Right}}: Dispersion in $x'_{targ}$ in mr/\% for particles from the same point at the IFP. Similar behavior is observed at other positions along the IFP. Both panels are shown on a linear scale.}
\label{fig:quadsondispersion}
\end{figure*}

Figure \ref{fig:quadsondispersion} shows the resulting dispersion of $x$ and $x'$ for particles with $x_{IFP}$ $=$ 0 $\pm$ 0.5 cm.
Simulations for other choices of $x_{IFP}$ lead to similar behaviour, though with an offset in $x_{targ}$. 
We see that the dispersion at the target is approximately 9.5 cm/\%, which is nearly 30 \% larger than the IFP dispersion. 
If we select a narrow region at the IFP experimentally, the position and width in $x$ of the beam spot at the target is a measure of the central momentum and width in momentum at the IFP. 
If different particles have different average momenta for the same IFP position, they are dispersed by 9.5 cm/\% at the target. 
If these particles differ in momentum by 1 \% at the IFP they would shift by nearly the full 10 cm GEM width at the target. 

For a beam arising from a point source, a 1-cm wide region at the IFP corresponds to a 0.14 \% momentum range, and generates an $x_{targ}$ range of a few cm. The first-order matrix elements presented, $( x_{IFP} / x_{prod} ) = -1.225$, along with the approximately 1.28 cm muon source size in $x$ and a dispersion of 7 cm/\% suggest that for muons a 1-cm wide slit at the IFP has a momentum range of about 1.28 cm $\times$ 1.225 / (7 cm/\%) $\approx$ 0.23 \%, which is about twice as large as for a point source, and which should make the beam spot about twice as large. The right column of Fig. \ref{fig:quadsondispersion} shows similar effects for the angle $x'$, with a dispersion in angle of approximately 25 mrad/\%, but the range of $\delta$ covered is larger for a fixed $x'$ than for a fixed $x$, so $x$ is a more sensitive quantity.

As a check of whether our conclusions are affected by too small a source distribution, the calculations were redone using the larger source size of the muon distribution beam at the IFP. The large source used $\pm$10.5 cm in $x$, $\pm$27 mrad in $x'$, $\pm$5 cm in $y$, $\pm$35 mrad in $y'$, and $\pm$2 \% in $\delta$, and again lacked correlations between parameters. We found the range of the correlations is larger -- there is more momentum acceptance leading to a broader beam spot -- but otherwise the correlations are unchanged.

To summarize, when taking measurements with a narrow collimator opening, the width of the beam spot in $x$ and $x'$ reflects the range of momentum of the particles within the collimator opening. Particles with a broader momentum range, for the same point at the IFP, will have a broader beam spot at the target. If different particles have different average momenta, the centers of their distributions will be shifted.
The dispersion is about 9 cm/\% in $x$ and 25 mrad/\% in $x'$. A 0.1 \% relative average momentum measurement requires determining the relative centers of the distributions to 9 mm or 2.5 mrad.

%% file: g4beamlinesim.tex
The G4beamline \cite{G4beamline} simulation software is specifically designed to simulate particles traveling down beam lines using the GEANT4 \cite{Agostinelli:2002hh} toolkit. 
G4beamline and TURTLE fundamentally differ in the way that they propagate particles through electromagnetic fields.  
TURTLE uses matrix algebra to act on a vector  $X=(x, x', y, y', \ell, \delta)$ that represents the particle's position and kinematics, with matrices that represent the electromagnetic fields.
In contrast, GEANT4 and therefore G4beamline, propagate particles through electromagnetic fields in short individual \enquote{steps}, accounting for the forces on the particles at each step.

The G4beamline description of the PiM1 channel was developed from the TURTLE description of the channel.
For a simple quadrupole in PiM1, it is possible to describe the geometry of the magnet and include the field strength in such a way that G4beamline can calculate the field. 
The ASM dipoles are more complicated than the models available in G4beamline, as the magnetic pole faces are curved. 
A field map was generated, though for a simpler dipole geometry than that used in TURTLE, for the simulation. 
While TURTLE and G4beamline should in principle produce consistent results, the use of both provides an estimate of the uncertainties in the calculations.

\subsection{Comparison of TURTLE and G4beamline Simulations}
\label{sec:simconparison}

\begin{figure*}[htb!]
\centering
	\includegraphics[width=\linewidth]{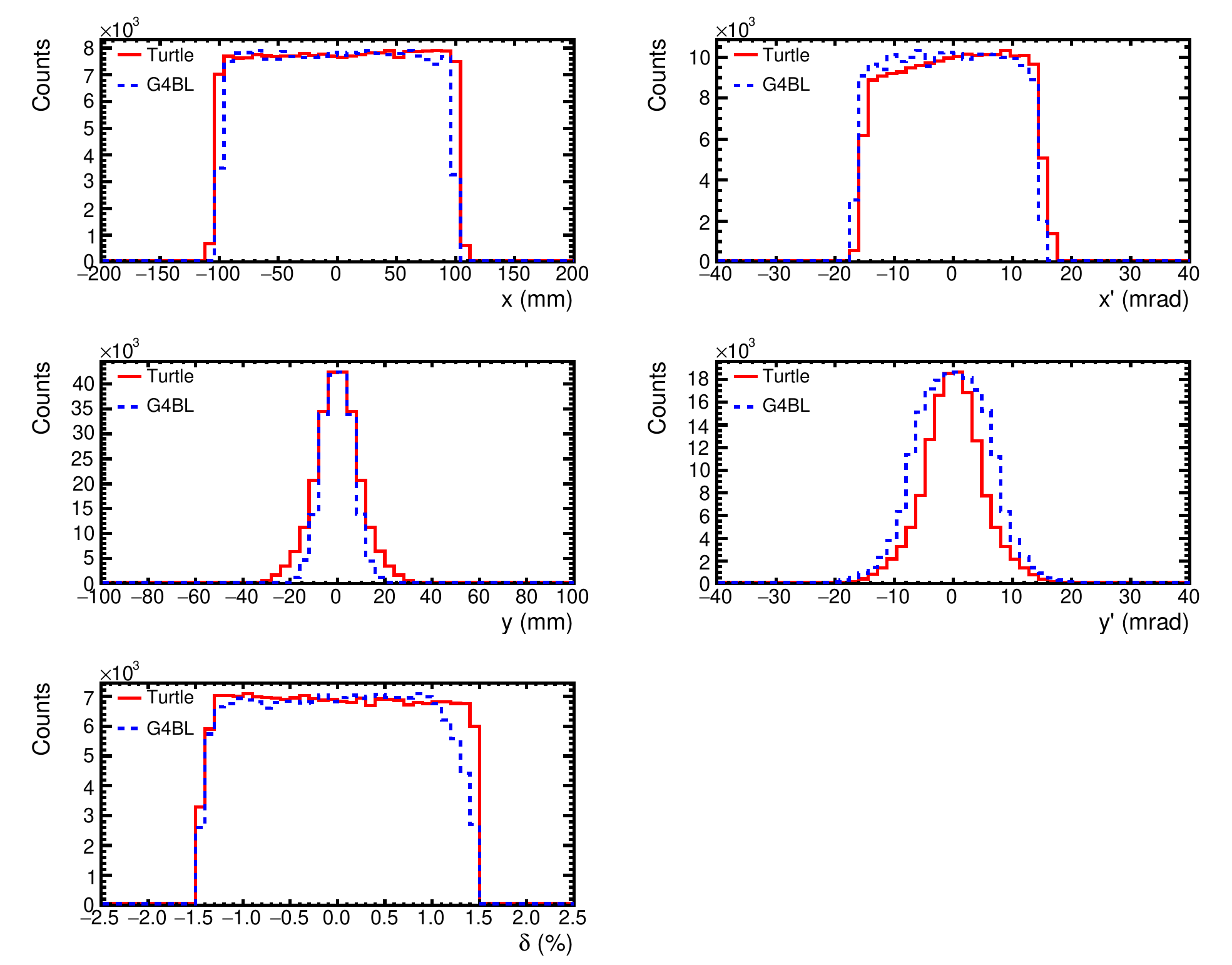}
	\caption[IFP distributions for a point source]{The distributions at the IFP using a point source for both TURTLE and G4beamline. TURTLE histograms are shown in red and G4beamline histograms, labeled as ``G4BL", are in blue. No energy loss material was included at the IFP.   \textbf{\textit{Upper Left:}}
$x$ distribution. \textbf{\textit{Upper Right:}} $x'$ distribution. \textbf{\textit{Middle Left:}} $y$ distribution.  \textbf{\textit{Middle Right:}} $y'$ distribution. \textbf{\textit{Lower Left:}} Relative Momentum, $\delta$, distribution.\label{fig:IFPpoint}}
\end{figure*}

\begin{figure*}[htb!]
\centering
	\includegraphics[width=\linewidth]{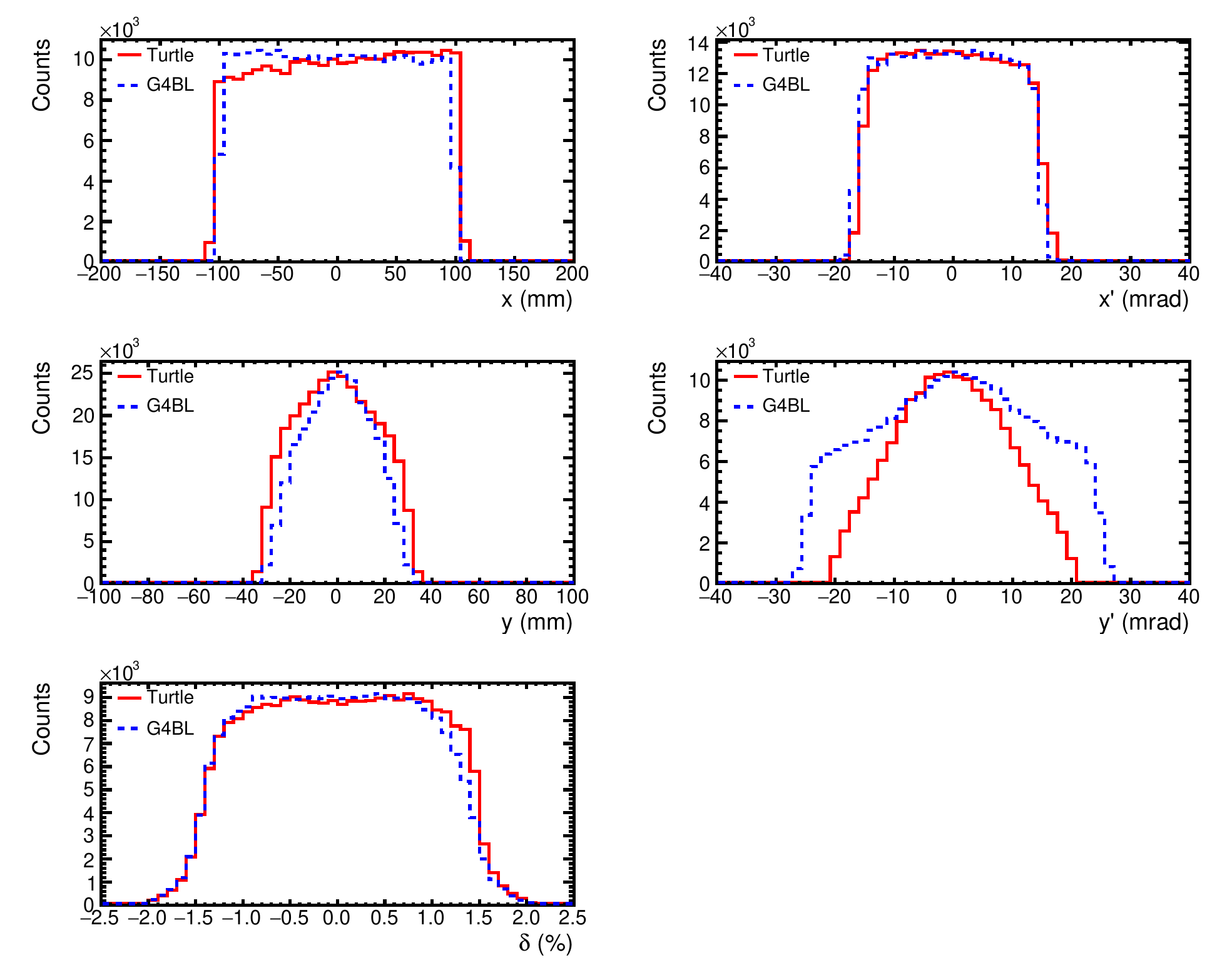}
	\caption[IFP distributions for a large source]{The distributions at the IFP using a muon source for both TURTLE and G4beamline. TURTLE histograms are shown in red and G4beamline histograms, labeled as ``G4BL", are in blue. No energy loss material was included at the IFP. \textbf{\textit{Upper Left:}}
$x$ distribution. \textbf{\textit{Upper Right:}} $x'$ distribution. \textbf{\textit{Middle Left:}} $y$ distribution.  \textbf{\textit{Middle Right:}} $y'$ distribution. \textbf{\textit{Lower Left:}} Relative Momentum, $\delta$, distribution.\label{fig:IFPlarge}}
\end{figure*}

\begin{figure*}[htb!]
\centering
	\includegraphics[width=\linewidth]{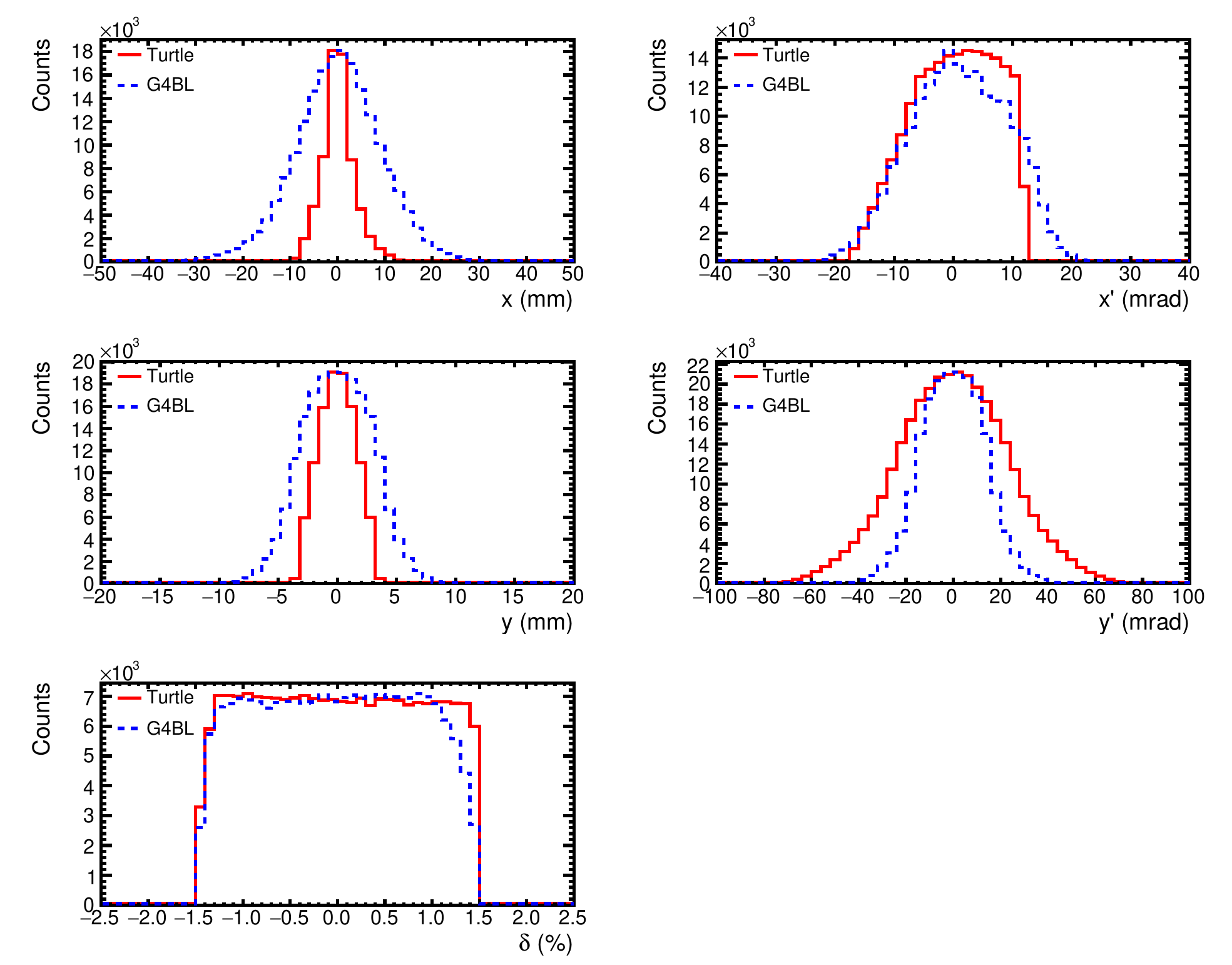}
	\caption[PiM1 distributions for a point source]{The distributions at the PiM1 target using a point source for both TURTLE and G4beamline. TURTLE histograms are shown in red and G4beamline histograms, labeled as ``G4BL", are in blue. No energy loss material was included at the IFP. \textbf{\textit{Upper Left:}}
$x$ distribution. \textbf{\textit{Upper Right:}} $x'$ distribution. \textbf{\textit{Middle Left:}} $y$ distribution.  \textbf{\textit{Middle Right:}} $y'$ distribution. \textbf{\textit{Lower Left:}} Relative Momentum, $\delta$, distribution.\label{fig:targetpoint}}
\end{figure*}

\begin{figure*}[htb!]
\centering
	\includegraphics[width=\linewidth]{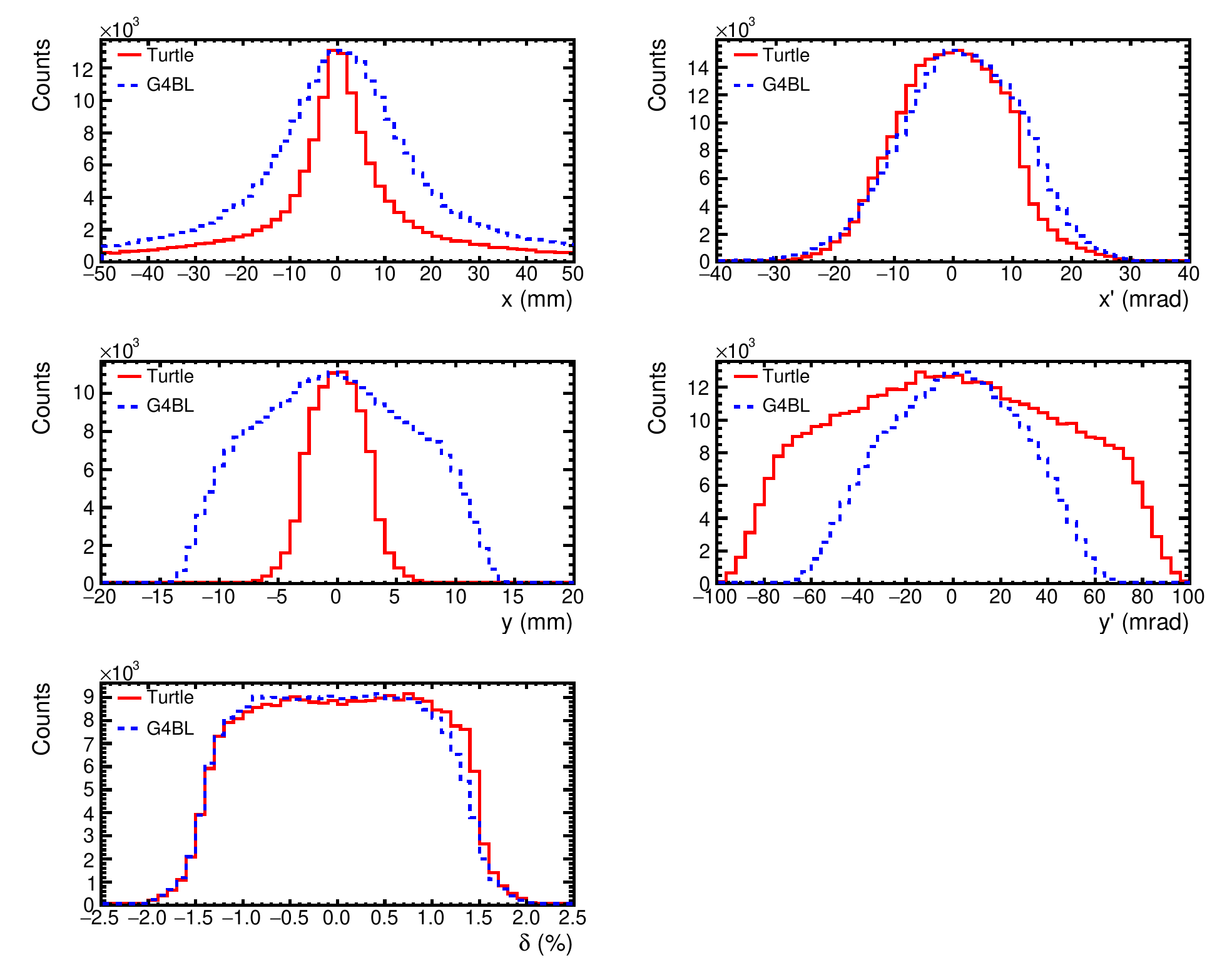}
	\caption[PiM1 distributions for a large source]{The distributions at the PiM1 target using a muon source for both TURTLE and G4beamline. TURTLE histograms are shown in red and G4beamline histograms, labeled as ``G4BL", are in blue. No energy loss material was included at the IFP.  \textbf{\textit{Upper Left:}}
$x$ distribution. \textbf{\textit{Upper Right:}} $x'$ distribution. \textbf{\textit{Middle Left:}} $y$ distribution.  \textbf{\textit{Middle Right:}} $y'$ distribution. \textbf{\textit{Lower Left:}} Relative Momentum, $\delta$, distribution. \label{fig:targetlarge}}
\end{figure*}

We now compare TURTLE and G4beamline at both the IFP and PiM1 target, for both point-like and muon sources, using the standard channel tune.
Here we consider the case of no energy loss along the beam line. This gives the cleanest comparison between both simulations, but is not sufficient for comparison to data.

Figures~\ref{fig:IFPpoint}$-$\ref{fig:targetlarge} show TURTLE and G4Beamline give generally similar results. The agreement is better at the IFP than at the target, and better for a point source than for an extended source, where details of the field descriptions would be expected to be more significant. G4beamline predicts distributions that are spatially smaller but broader in angle than TURTLE at the IFP. The reverse is true at the target. TURTLE predicts slightly more momentum acceptance at both IFP and target.

%% file: comp.tex
\label{sec:comp}

We now compare the results of the G4beamline simulations to several measurements of the beam properties, to test the quality of the simulations.
In order to have quantitative agreement between the beam line simulation and data, it is necessary to consider the effects of multiple scattering in the beam, and energy loss and multiple scattering, especially at the IFP, but also to a lesser degree after the beam exits the channel and traverses air before striking the detectors and target.
We compare only the G4beamline simulation to the data, due to the superior implementation of material effects in the G4beamline simulation, which is based on Geant4.

\subsection{Momentum Dispersion at the IFP}
\label{sec:disp}
\begin{figure}
    \centering
    \includegraphics[width=\linewidth]{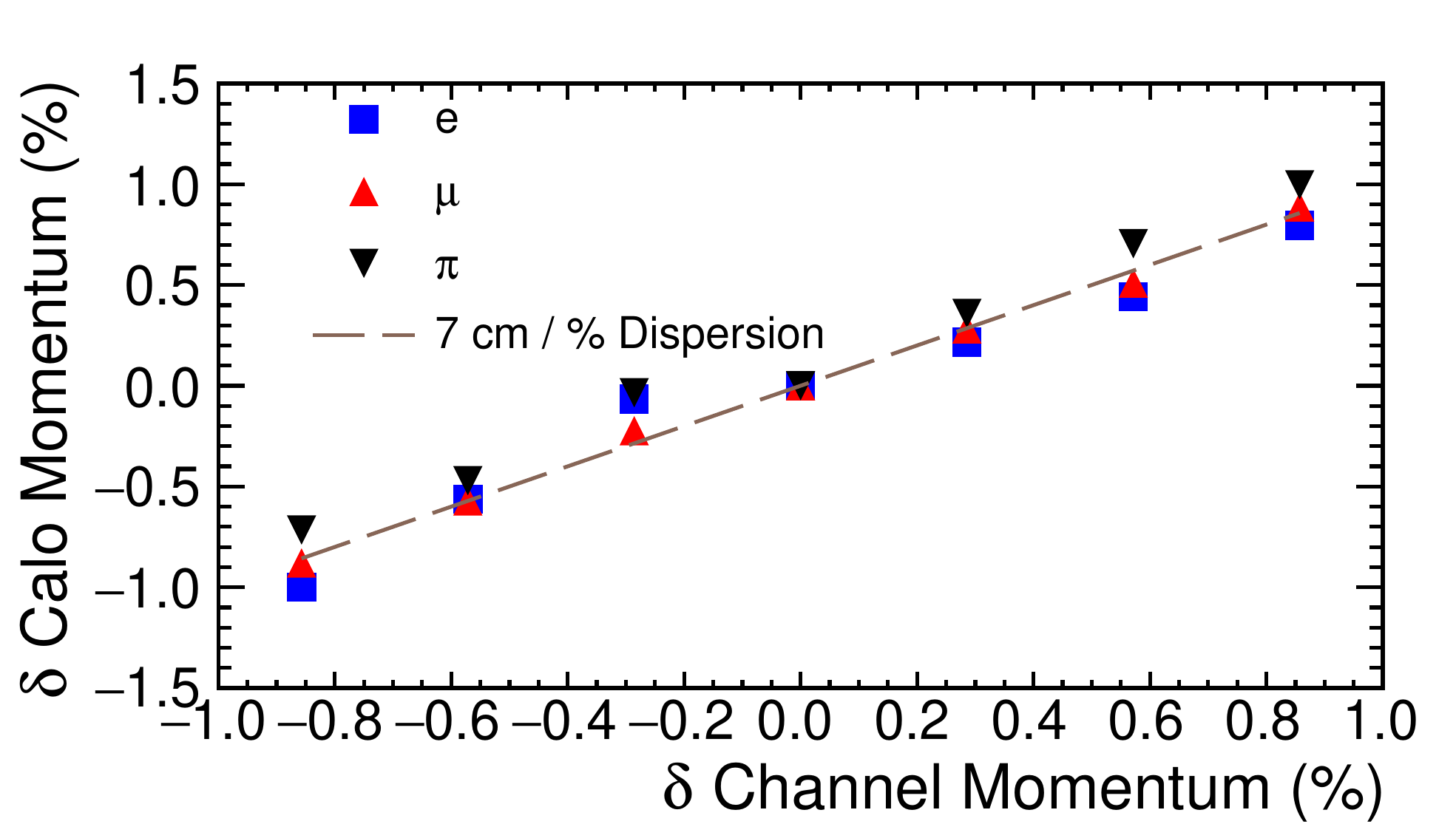}
    \caption{A comparison between the energy deposited in the beam line Calorimeter and the IFP collimator position relative to the central IFP collimator position, for a channel setting of +160 MeV/c. The brown dashed line, normalized to the central 0 \% point, shows how the energy deposition is expected to vary with the expected dispersion of 7 cm/\%. Only statistical error bars are shown.}
    \label{fig:edep_v_coll}
\end{figure}

Figure~\ref{fig:edep_v_coll} shows the measured dispersion of the channel at the IFP to be 7 cm/\%, in good agreement with the simulations.
The total energy of the electrons in the beam were measured by a calorimeter in PiM1.
Normalizing the expected dispersion curve to the light output of the calorimeter at the central momentum, at a collimator position of 0 cm, yields energy depositions that all agree with expectations to approximately 0.16 \%.
This is excellent agreement given the calorimeter resolution at the central beam momentum of 160 MeV/$c$ is about 14 \%.
An estimate of the energy loss is included by normalizing the calorimeter relative momentum to 0 for the central collimator position.

\subsection{Collimator Scan}
\label{sec:col_scan}

\begin{figure}
    \centering
    \includegraphics[trim= 0 0 0 0,clip,width=\linewidth]{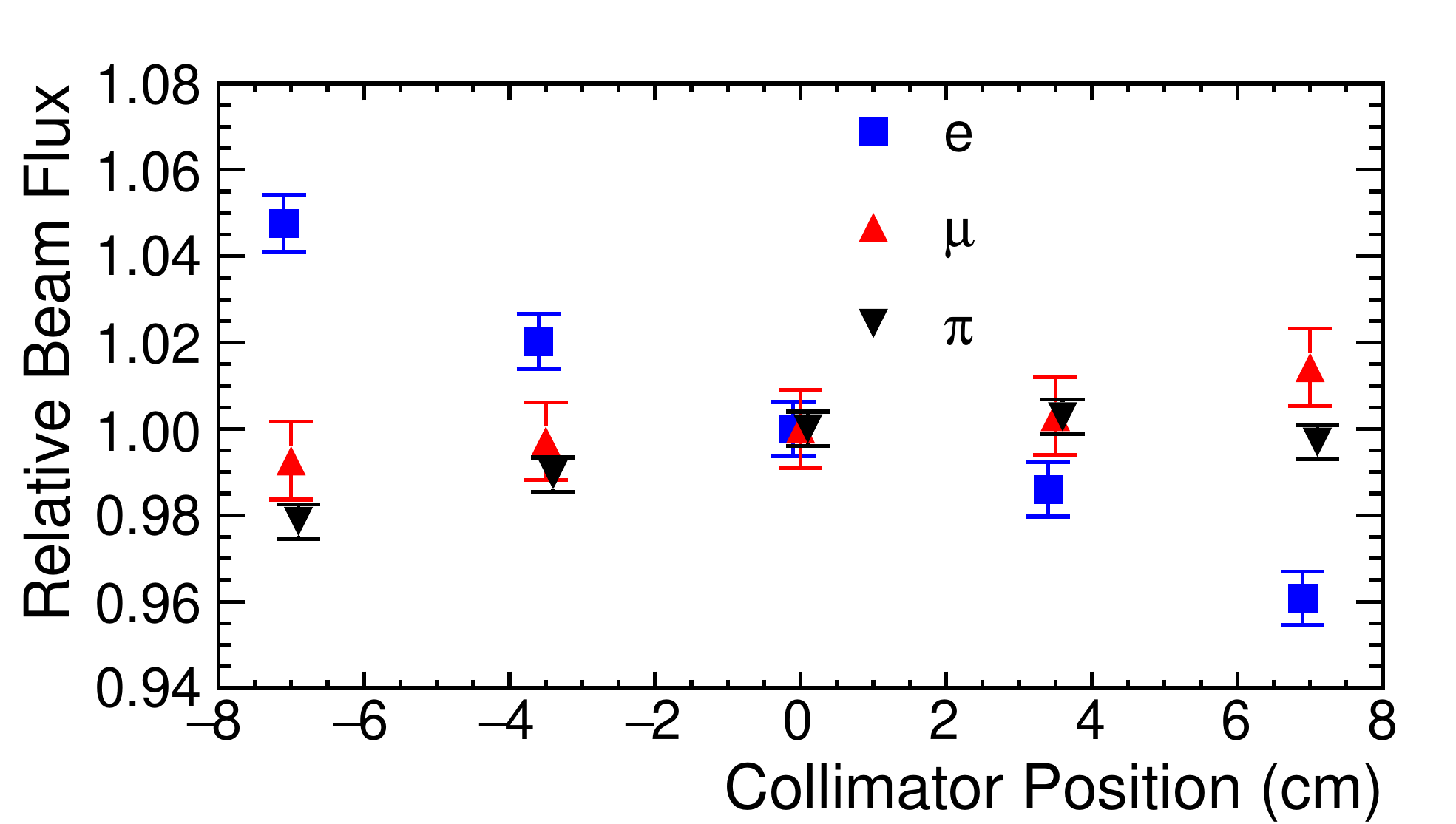}
    \caption{Beam flux at the PiM1 target as a function of IFP collimator position. Note that a $\pm$10 \% change in flux across the IFP corresponds to a shift in the average momentum of 0.03 \%. Points are horizontally offset for plotting. Uncertainties shown are statistical.}
    \label{fig:collscanflux_+160}
\end{figure}

\begin{figure}
    \centering
    \includegraphics[width=\linewidth]{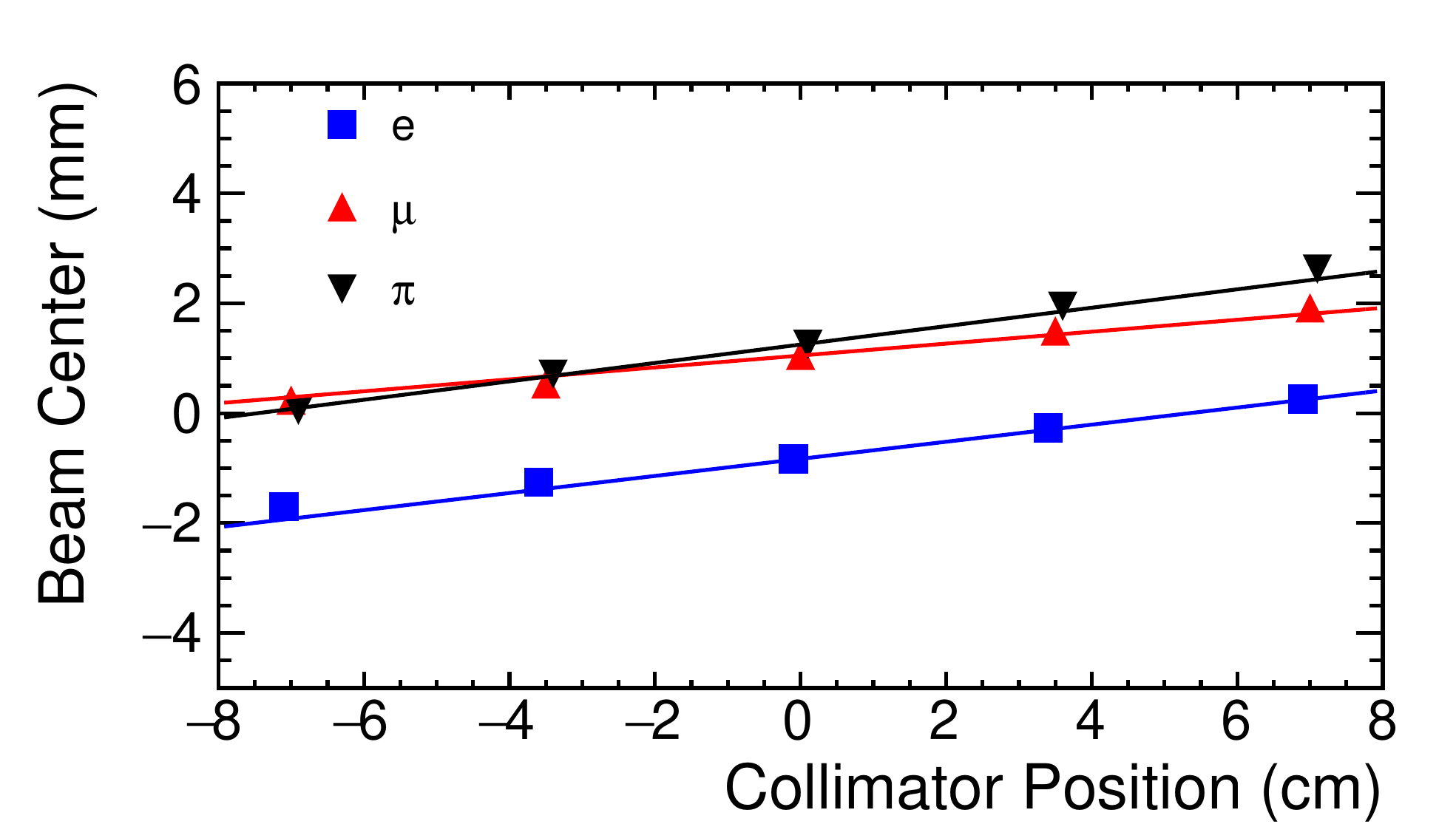}
    \caption{Beam spot position at the PiM1 target as a function of IFP collimator position. Data were taken at +160 MeV/$c$. The $\pm$7 cm range of collimator position  corresponds to $\pm$1 \% in momentum. The lines show calculations from G4beamline for each particle species. Note that the positions of different species are the same to within approximately 2 mm at each collimator setting. Data are plotted slightly horizontally offset for better visibility.}
    \label{fig:collscanpos_+160}
\end{figure}

A crucial issue for MUSE is knowledge of the absolute beam momentum, and the relative momentum of electrons and muons for comparisons between the two particle species.
The beam line central momentum is set with an accuracy of $10^{-4}$ by the dipole magnet power supplies, and can be checked with the accelerator Radio Frequency (RF) time measurements.
Here we consider how the simulations impact our knowledge of the momentum distribution of the beam and the relative momentum of particles.

The data and simulations show that the beam dispersion at the IFP is 7 cm/\%.
The momentum bite of the beam can be up to $\pm$1.5 \%.
MUSE operates with FS11 jaw and FS13 IFP collimator settings that limit the channel angle and momentum acceptance to reduce the flux to 3.5 MHz.
However, MUSE generally requires a 0.2 \% knowledge of the beam momentum -- see Appendix \ref{sec:mom_req} -- and the momentum profile needs to be determined for input to the experimental analysis.
To determine the momentum profile and average, collimator scans are performed, with a narrow collimator opening scanned across the IFP to determine the flux vs momentum.
Figure~\ref{fig:collscanflux_+160} shows such a collimator scan performed at +160 MeV/$c$.
For all particle species, the flux has a small, approximately linear, variation across the momentum acceptance.
We note that for a $\pm$1 \% momentum bite, a $\pm$10 \% flux change across the acceptance corresponds to a change in the average momentum of 0.03 \%, which is a small correction in comparison with the 0.2 \% requirement.

The collimator scan can also be used to precisely confirm the relative momenta of the different particle species.
This is determined from the beam spot positions at the target, shown in Fig.~\ref{fig:collscanpos_+160}.
The different species of beam particles are brought to a focus to within about 2 mm at the target position, for a given collimator setting, but as the collimator position is moved across the acceptance the beam spot positions all move by about 2 mm.

The former observation can be used to set a limit on whether the momenta of the particles are the same at the target.
At the IFP, the particles have to have the same average momenta at the same position, although the resolution is worse for muons due to the extended source.
This is because they have passed through a magnetic system, which selects momenta, but no materials.
Any change in position at the PiM1 target reflects different energy losses in the IFP region.
The dispersion from the IFP to the PiM1 target is given by the matrix element 
$\left< x_{targ} | \delta_{IFP} \right>$ = 9.5 cm/\%.
A 2-mm offset then corresponds to a 0.02 \% momentum difference, which is small compared to the momentum requirement of 0.2 \%.

The observation that all the peaks shift together across the acceptance by about 2 mm roughly corresponds to the $\left< x_{targ} | \delta_{prod} \right>$ matrix element of Table~\ref{tab:1storderme}: particles from the source with the same trajectory but different momenta have a dispersion at the target of 1.4 mm/\%, corresponding to a 2.8 mm shift. The full G4Beamline calculation reduces the dispersion to the 1 mm/\% shown in Fig.~\ref{fig:collscanpos_+160}.

\subsection{Shape of Particle Distributions at the Target}

\begin{figure*}
    \centering
    \includegraphics[trim=0 0 00 0,clip,width=\linewidth]{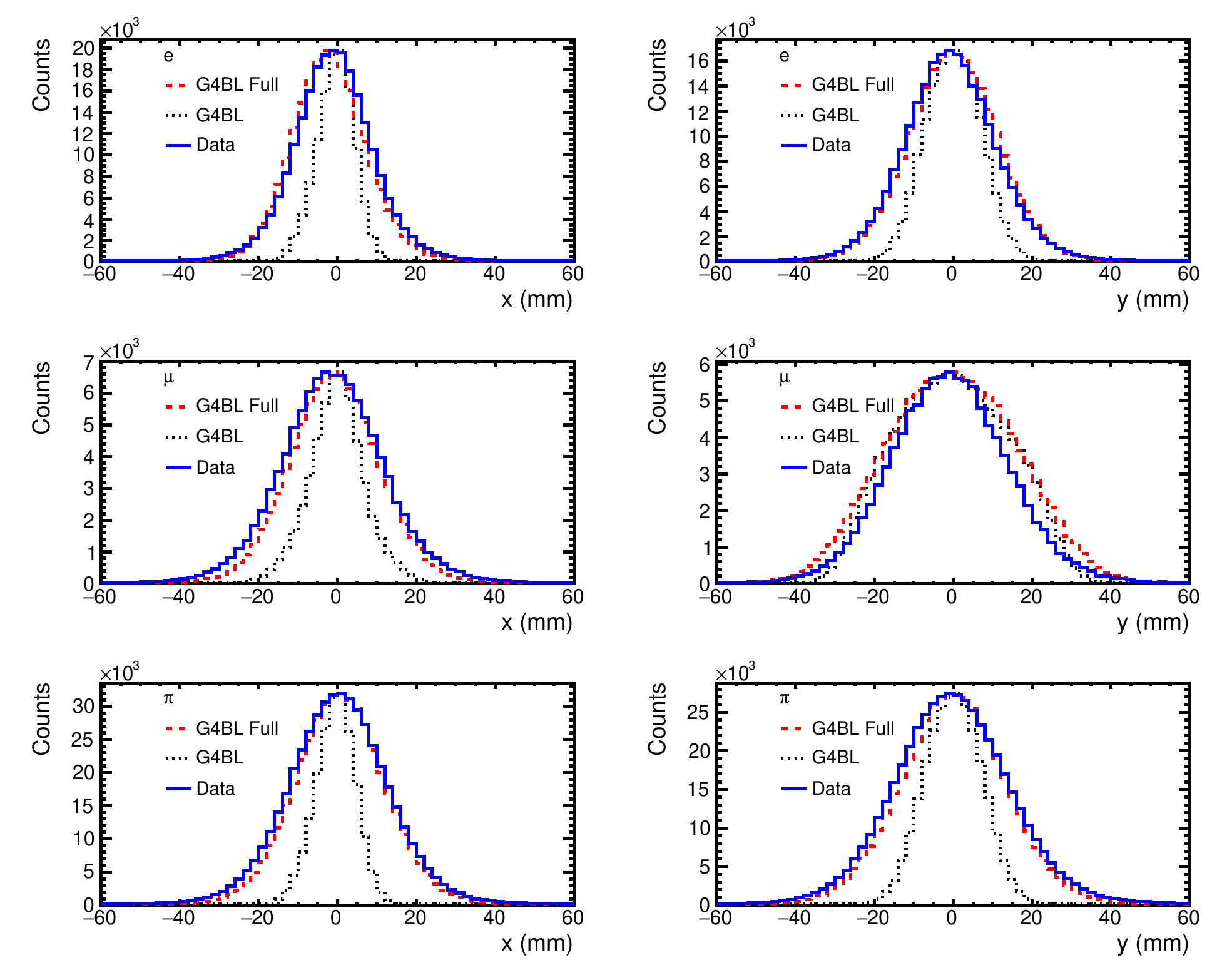}
    \caption{Comparison of G4beamline simulations and data at the MUSE target. The ``G4BL" simulation does not include energy loss, but the ``G4BL Full" does. \textbf{\textit{Left Column:}} horizontal distributions. \textbf{\textit{Right Column:}} vertical distributions.  \textbf{\textit{Top Row:}} electron distributions. \textbf{\textit{Middle Row:}} muon distributions. \textbf{\textit{Bottom Row:}} pion distributions. }
    \label{fig:G4_Data_Comp}
\end{figure*}

The beam spot at the MUSE target was simulated using both G4beamline and TURTLE. 
Here we only compare to G4beamline, including the effects of materials in the beam line, due to the superior energy loss and multiple scattering descriptions in G4beamline. 

Figure~\ref{fig:G4_Data_Comp} compares G4beamline simulated with measured particle distributions. We also include the ``simple" G4beamline simulation without energy loss, to show the necessity of including such calculations in the simulation.

In the data, particle times measured using the BH scintillator detector and the accelerator RF signal allow the different particle species to be identified.
The $x$ and $y$ distributions in the simulation are determined at the MUSE target position, while the distributions in the data are from the tracks measured by the beam line GEM chamber telescope projected to the target position.
The measured distributions are well reproduced aside from the longer tails seen in the data.
The longer tails in the data arise mainly from noise in the GEMs which cause clusters to be misidentified, leading to incorrect tracks.
These are preliminary tracking results, to which precise geometric alignment has not been applied.

\subsection{RF Timing Measurements}
\label{sec:rf_time}

\begin{figure*}
    \centering
    \includegraphics[width=\linewidth]{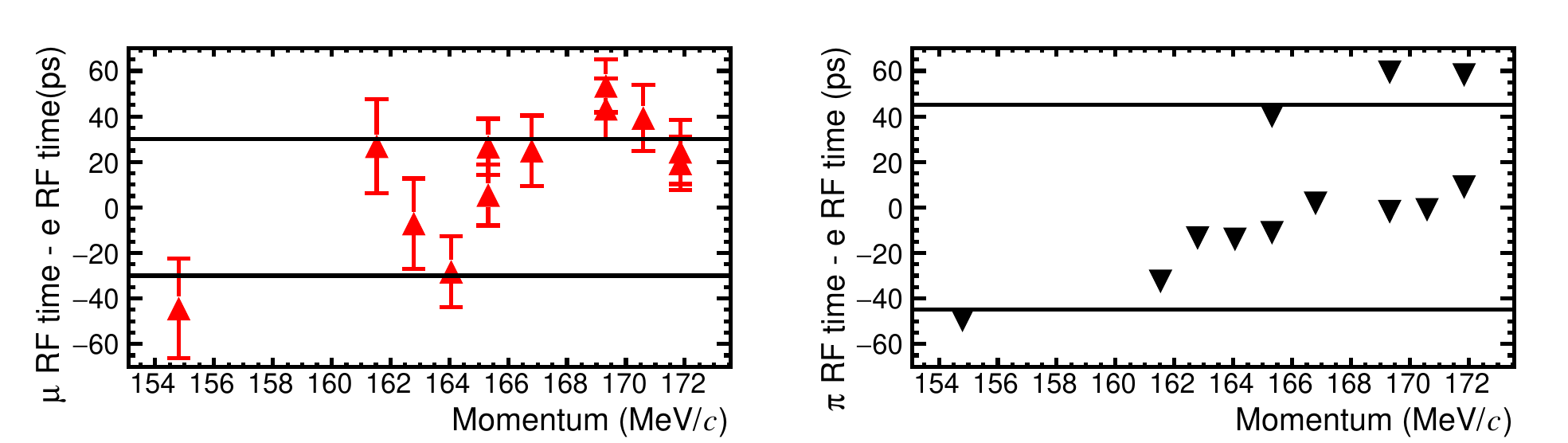}
    \caption{\textbf{\textit{Left:}} $\mu$ - $e$ RF distribution vs. momentum. The solid black lines indicate a $\pm$0.1 \% change in relative momentum. \textbf{\textit{Right:}} $\pi$ - $e$ RF distribution vs. momentum. The solid black lines indicate a $\pm$0.1 \% change in relative momentum. For both panels only statistical uncertainties are shown.}
    \label{fig:mue_pie_RF}
\end{figure*}

The RF times measured by the BH allow for identification of particle species,
since the particles arrive at the detector at a fixed phase relative to the accelerator RF signal. 
The accelerator RF 50.63 MHz signal is digitized and fed into the data acquisition system. 
The difference between RF and BH signal times determines particle type.
As the electrons have \textbeta~$\approx 1$ for all MUSE momenta, their RF position is the same for all momenta -- the $e$ RF peak can be used as a reference for the other particle species.
The difference in RF peak position between, e.g., the pions and electrons allows the absolute momentum to be determined if the path length is precisely known, and the momentum stability to be determined even if the path length is not precisely known. 
At MUSE momenta, a 0.1 \% shift in beam momentum for muons (pions) leads to an RF time shift of approximately 30 (45) ps. 
An example of using the RF time to determine momenta is shown in  Fig.~\ref{fig:mue_pie_RF}, where the muon-electron and pion-electron RF time difference was analyzed to determine the muon and pion momenta.
The channel length was determined from survey information of the magnets and the BH detector, assuming particles travel along a circular arc through each dipole magnet.
A correction was applied for energy loss of the pions and muons in the materials at the IFP.
The black lines in Fig.~\ref{fig:mue_pie_RF} indicate the timing shift corresponding to 0.1 \% momentum shift. 
It can be seen that the RF peaks are consistent with the time expected from the set channel momenta to within approximately $\pm$0.1 \%.
It is important to note that this momentum stability check is performed continuously during production data taking.

\subsection{RF Timing Measurements and Collimator Scan to Verify Dispersion}

\begin{figure*}[htb!]
    \includegraphics[width=\textwidth]{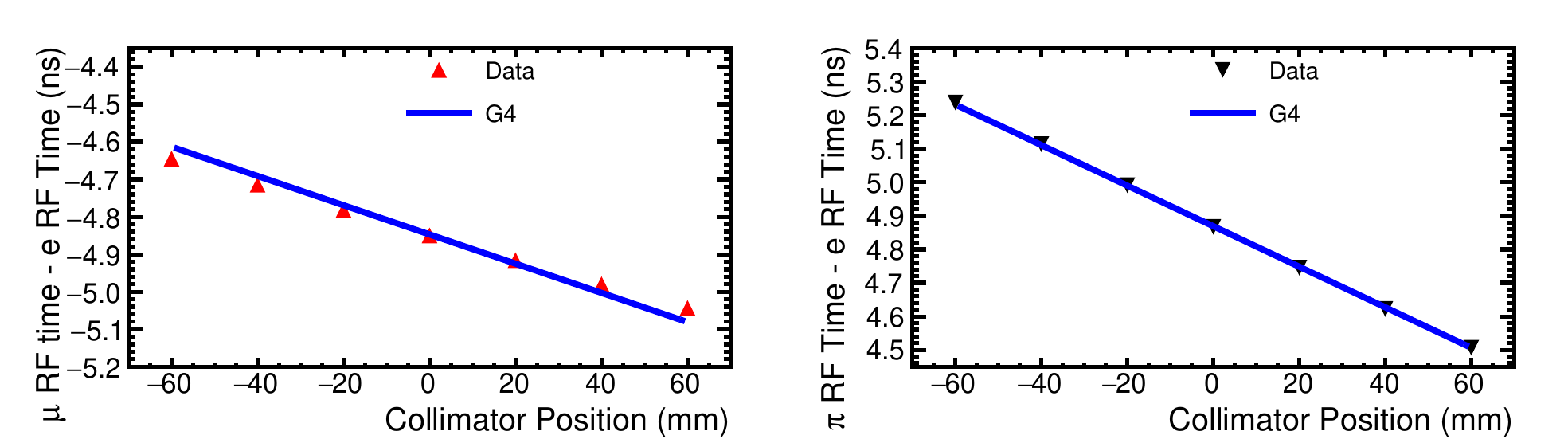}
    \caption{Verification of the dispersion at the IFP via a comparison of measured shifts in RF peak position and the G4beamline simulation. Only statistical uncertainties are shown. \textbf{\textit{Left:}} $\mu - e$ RF distribution vs. collimator position. \textbf{\textit{Right:}} $\pi - e$ RF distribution vs. collimator position. }
    \label{fig:rftime_collscan}
\end{figure*}

The momentum dispersion at the IFP can also be verified with the muon-electron and pion-electron RF time differences when a collimator scan is performed. 
As described in Secs.~\ref{sec:disp}, \ref{sec:col_scan}, and \ref{sec:rf_time}, there is a 7 cm/\% dispersion at the IFP, and the IFP collimator selects the relative momentum bite.
We apply the same RF time techniques as in Sec.~\ref{sec:rf_time}, except that when varying the central momentum, the path length stays constant, while when varying the IFP collimator setting, the path length changes.
Figure~\ref{fig:rftime_collscan} shows a comparison of the measured RF time differences to a calculation from G4beamline.
The agreement between data and simulation is good, to within approximately 30 ps or 0.1 \%, for only a momentum offset, or about 7 mm, for only a path length offset. 
We would expect any discrepancies with the path length calculation to be larger for the muons than for the pions, due to the larger source size, and for edge rather than central momentum bites, due to the need to better describe the magnetic fields. 
The agreement of data and simulation for these RF time measurements, the calorimeter measurements of Sec.~\ref{sec:disp}, and the target dispersion measurements of  Sec.~\ref{sec:col_scan} reinforce that the dispersion at the IFP is known and the relative momenta of the particles is well understood.

%% file: Conclusion.tex
The PiM1 channel was developed to generate a high-precision pion beam line. For this case, it has  well-studied properties, which we presented.
The properties of the electron and muon beams also present in the channel had not been well established previously, and might be significantly different due to the different production mechanisms for these secondary particles.
We presented simulations of electron and muon production at the M target. We concluded that the electron production happens over essentially the same spatial region as the pion production, so the electron and pion beams have the same properties, aside from differences in the shape of the momentum distribution and differences induced by multiple scattering and energy loss at the IFP.
Muon production happens over a wider spatial region, so the muon beam has tails in the spatial, angle, and momentum distributions, as well as poorer resolution.

We compared the G4beamline and TURTLE simulations of the transport of the same source distributions to the channel IFP and scattering target positions.
The simulations are overall fairly similar.
The TURTLE simulation has a more precise description of the dipole fields, while the G4beamline simulation better describes the effects of the IFP air and window material on the particle beams.

We presented measurements of the beam properties, compared to expectations from known properties and the simulations.
The momentum dispersion at the IFP, beam position, and relative beam flux was studied using the MUSE experimental apparatus and collimator.
The known 7 cm/\% dispersion, reproduced by the simulations, was confirmed in our measurements using the beam calorimeter.
The beam flux measurements showed that the average momentum of particles passing through the channel agrees with the central set momentum to within 0.03 \%.
The beam spot for each particle type at the PiM1 target was seen to move in accordance with the 1.4 mm/\% dispersion from the M target.
The positions of the different particle species were observed to be consistent at the roughly 2-mm level, indicating their momenta are consistent to within approximately 0.02 \%.
RF time measurements of particles propagating through the channel showed approximately 0.1 \% agreement with the set momentum.

We conclude that the muon and electron beams have quite similar properties to the pion beam and to each other, so that the PiM1 channel can be used for high-precision muon and electron scattering.

%% file: Momentum_requirements.tex
\begin{figure}
\centering
\includegraphics[width=\linewidth]{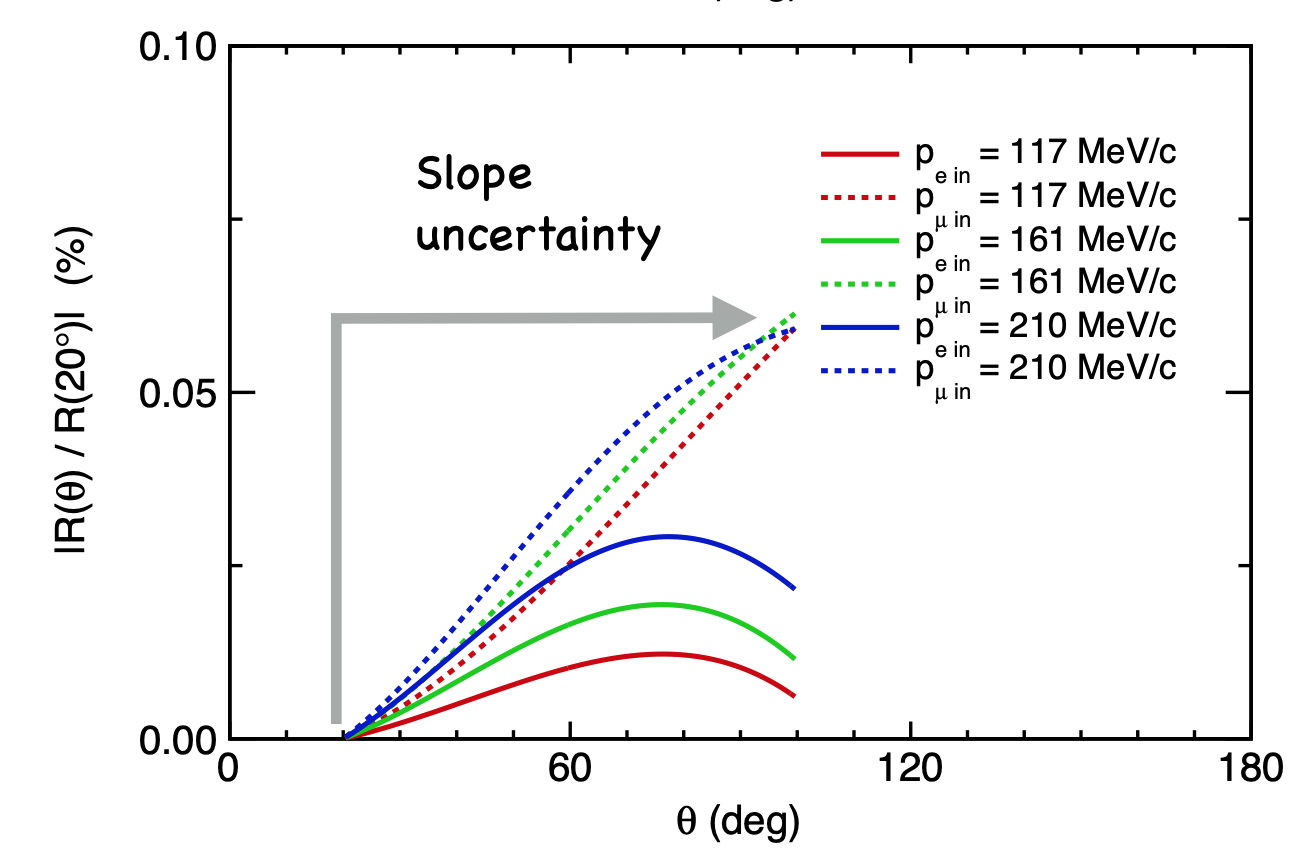}
\caption[Effect of beam momentum offset on the cross section]{Change in cross section in percent for a 0.1 \% change in the beam momentum over the MUSE angle range. Using the Kelly form factor parameterization \citep{Kelly:2004hm}.}
\label{fig:energysensitivity}
\end{figure}

 In order to precisely extract the proton charge radius, MUSE must understand and precisely measure the momentum of the beam line. 
 The technical design for MUSE specifies that the beam momentum must be known to $0.2\%$ for muons and $0.3\% - 0.8\%$ for electrons in order to achieve the required 0.1\% (0.05\%) uncertainty on the respective cross sections ($G_E$).  
 Figure \ref{fig:energysensitivity} shows the effect of momentum offsets on the cross section. 

%% file: M_Target.tex
The beam line was originally designed to host a pion beam, with pions being generated by the $pC \rightarrow \pi X$ reaction. Historically, as QCD was being developed in the 1970’s, several pion factories were constructed. Cyclotrons were developed at TRIUMF and PSI, and a linear accelerator at Los Alamos. These accelerators were all designed primarily to provide high pion flux beam lines in order to provide a more fundamental understanding of the strong force over a range of energies. As pions scatter with targets via the strong interaction, electron and muon contamination of the beams and scattering interactions via electromagnetic scattering were generally a minor issue for pion scattering measurements that could be suppressed by a variety of techniques. Furthermore, the pion beam lines were generally unsuitable for muon studies due to an overwhelming pion background, so the properties of the background muons and electrons in pion beam lines were generally not well established. The graphite M Target was constructed with pion beam properties in mind. The properties of the M Target are summarized in Table \ref{table:MTarget}.
\begin{table}[htb!]
\caption[Properties of the M target]{Properties of the M target \cite{MTarget}.}
\label{table:MTarget}
\begin{tabular}{ c@{\hskip 0.5in}c } 
 \hline
 \hline
M Target Property & Value  \\ 
\hline
Mean Diameter & $320$ mm \\
 Target Thickness & $5.2$ mm \\
 Target Width & $20$ mm \\
 Graphite Density& $1.9$ g/cm$^{3}$ \\
 Beam Loss & $1.6$~\% \\
 Power Deposition & $2.4$ kW/mA \\
  Operating Temperature & $1100$ k \\
 Irradiation Damage Rate & $0.12$ dpa/Ah \\
 Rotation Speed & $60$ rpm  \\
 \hline
 \hline
\end{tabular}

\end{table}